%% file: arxiv.tex
\documentclass[submission, Phys]{SciPost}
\usepackage[utf8]{inputenc}
\usepackage[T1]{fontenc}
\usepackage{booktabs}
\usepackage{tabularx}
\usepackage{graphicx}
\usepackage{xspace}
\usepackage{multirow}
\usepackage{amssymb,amsmath}
\usepackage{mathtools}
\usepackage{cleveref}
\usepackage{xstring}
\usepackage{mathrsfs}
\usepackage{textcomp}
\usepackage{listings}
\usepackage{accsupp}
\usepackage[most]{tcolorbox}
\usepackage[inline]{enumitem}
\usepackage{pmboxdraw}
\usepackage{paracol}
\usepackage{noindentafter}

\newcommand{\codenum}{\ttfamily\tiny\color{gray}}

\tcbset{bbox/.style={breakable,colback=white,colframe=black,
    boxrule=0.4pt,boxsep=0pt,left=2pt,right=4pt}}
\tcbset{nbbox/.style={colback=white,colframe=black,
    boxrule=0.4pt,boxsep=0pt,left=2pt,right=4pt}}
\BeforeBeginEnvironment{lstlisting}{\begin{tcolorbox}[bbox]\vspace{-10pt}}
\AfterEndEnvironment{lstlisting}{\vspace{-10pt}\end{tcolorbox}}
\NewDocumentCommand{\code}{O{}O{bbox}m}{\begin{tcolorbox}[#2]%
    \vspace{-10pt}\lstinputlisting[#1]{#3}\vspace{-10pt}\end{tcolorbox}%
  \noindent}

\NoIndentAfterEnv{lstlisting}

\crefname{line}{line}{lines}
\Crefname{line}{Line}{Lines}

\DeclareRobustCommand{\ccite}[1]{\IfSubStr{#1}{,}{refs.~}{ref.~}\cite{#1}}
\DeclareRobustCommand{\Ccite}[1]{\IfSubStr{#1}{,}{Refs.~}{Ref.~}\cite{#1}}

\newcommand{\eg}{\textit{e.g.},\xspace}

\newcommand{\fortran}{\vb{FORTRAN}\xspace}
\newcommand{\python}{\vb{Python}\xspace}
\newcommand{\cpp}{\vb{C++}\xspace}
\newcommand{\posix}{\vb{POSIX}\xspace}
\newcommand{\xml}{\vb{XML}\xspace}

\newcommand{\scb}[1]{\texorpdfstring{\protect\scalebox{0.8}{#1}}{#1}}
\newcommand{\scm}[1]{\ensuremath{\text{#1}}}
\newcommand{\mlhad}{\scm{ML\scb{HAD}}\xspace}
\newcommand{\pythia}{\scm{P\scb{YTHIA}}\xspace}

\newcommand{\contrib}{\scm{P\scb{YTHIA}}-\scm{C\scb{ONTRIB}}\xspace}
\newcommand{\fastjet}{\scm{F\scb{AST}}\scm{J\scb{ET}}\xspace}

\newcommand{\rivet}{\scm{R\scb{IVET}}\xspace}

\newcommand{\cmake}{\scm{C\scb{MAKE}}\xspace}
\newcommand{\doxygen}{\scm{D\scb{OXYGEN}}\xspace}
\newcommand{\dox}{\scm{D\scb{OX.}}\xspace}

\hypersetup{colorlinks, linkcolor={red!50!black}, citecolor={blue!50!black},
  urlcolor={blue!80!black}}

\definecolor{comments}{rgb}{0.0, 0.4, 0.0}
\lstdefinestyle{shell}{
  numbers=none,
  xleftmargin=2pt,
  breakatwhitespace=true,
  literate={},
  prebreak={\space\textbackslash},
}

\ifdefined\pdfgentounicode
  \input{glyphtounicode}
\fi
\newcommand{\nocopy}[1]{\BeginAccSupp{ActualText={}}#1\EndAccSupp{}}

\newcommand{\vb}[1]{\lstinline[alsoother={_}]!#1!}
\providecommand{\tightlist}{\setlength{\itemsep}{0pt}\setlength{\parskip}{0pt}}

\newcommand{\reportnumber}{FERMILAB-PUB-26-0739-CSAID, MCNET-26-30}
\usepackage[angle=0,scale=1,color=black,firstpage=true,opacity=1]{background}
\SetBgContents{\footnotesize\textsf{\reportnumber}}
\SetBgPosition{current page.north east}
\SetBgHshift{-2in}
\SetBgVshift{-0.5in}

\begin{document}

\begin{center}
  \textbf{\Large Pythia-Contrib: a modular contribution framework for the Pythia 8 event generator}
\end{center}

\begin{center}
Beno\^ it Assi\textsuperscript{2,3$\spadesuit$},
Christian Bierlich\textsuperscript{4$\clubsuit$},
Rikab Gambhir\textsuperscript{1$\diamond$},
Phil Ilten\textsuperscript{1$\dagger$},
Tony Menzo\textsuperscript{5,6$\star$},
Stephen Mrenna\textsuperscript{7$\maltese$},
Manuel Szewc\textsuperscript{1,8$\parallel$},
Michael K. Wilkinson\textsuperscript{1,9$\perp$},
and Jure Zupan\textsuperscript{1$\mathsection$}
\end{center}

\begin{center}
\textsuperscript{\bf 1} Department of Physics, University of Cincinnati, Cincinnati, Ohio 45221, USA \\
\textsuperscript{\bf 2} Perimeter Institute for Theoretical Physics, Waterloo, N2L 2Y5, Canada \\
\textsuperscript{\bf 3} Department of Physics, University of Toronto, Toronto, ON, M5S 1A7, Canada \\
\textsuperscript{\bf 4} Department of Physics, Lund University, Box 118, SE-221 00 Lund, Sweden \\
\textsuperscript{\bf 5} Department of Physics, University of Alabama, Tuscaloosa, AL 35487, USA \\
\textsuperscript{\bf 6} Theory Division, Fermilab, Batavia, Illinois, USA \\
\textsuperscript{\bf 7} Scientific Computing Division, Fermilab, Batavia, Illinois, USA \\
\textbf{\textsuperscript{8}} International Center for Advanced Studies (ICAS), ICIFI and ECyT-UNSAM, 25 de Mayo y Francia, (1650) San Mart\'{i}n, Buenos Aires, Argentina \\
\textsuperscript{\bf 9} Department of Physics,
University of Oklahoma,
Norman, OK 73019, USA\\
${}^\spadesuit${\small \sf bassi@perimeterinstitute.ca},
${}^\clubsuit${\small \sf christian.bierlich@hep.lu.se},
${}^\diamond${\small \sf gambhirb@ucmail.uc.edu},
${}^\dagger${\small \sf philten@cern.ch},
${}^\star${\small \sf 
amenzo@ua.edu},
${}^\maltese${\small \sf mrenna@fnal.gov},
${}^\parallel${\small \sf mszewc@unsam.edu.ar},
${}^\perp${\small \sf michael.k.wilkinson@ou.edu},
${}^\mathsection${\small \sf zupanje@ucmail.uc.edu}
\end{center}

\begin{center}
\includegraphics[width=0.4\textwidth]{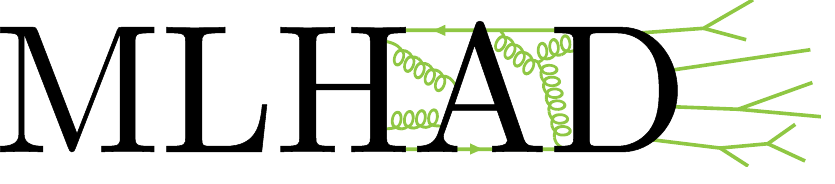}
\end{center}

\section*{Abstract}
\pythia is a general-purpose Monte Carlo event generator used in particle, astroparticle, and nuclear physics, by theorists and experimentalists, to simulate particle collisions.
Because of the size and diversity of the user community, it is challenging to address all of the needs for new and varied physics models.
This paper describes the \contrib system, which allows for third-party developers to contribute code to \pythia without introducing external software dependencies to the \pythia code base.

\vspace{10pt}
\noindent\rule{\textwidth}{1pt}
\tableofcontents\thispagestyle{fancy}
\noindent\rule{\textwidth}{1pt}
\vspace{10pt}

\section{Introduction}

\pythia is a general-purpose Monte-Carlo event generator (MCEG) which simulates the collisions of high-energy particles, with code originally written in \fortran\cite{Bengtsson:1982jr,Bengtsson:1984yx,Bengtsson:1987kr,Sjostrand:1993yb,Sjostrand:2000wi,Sjostrand:2001yu,Sjostrand:2003wg,Sjostrand:2006za}, later rewritten into \cpp\cite{Sjostrand:2007gs,Sjostrand:2014zea,Bierlich:2022pfr}.
A \python interface to \pythia has also been available since version \vb{8.219}\cite{PythiaPyPi}, first via \textsc{Swig}\cite{Swig}, and then \textsc{PyBind11}\cite{PyBind11}.
Throughout its more than 40-year history\cite{Sjostrand:2019zhc}, \pythia has remained open source with core code free of external software dependencies.
This policy has contributed to the extended success of \pythia, as users can quickly and easily begin running \pythia without significant overhead.
Indeed, \pythia has been adopted extensively within the particle-physics community; \pythia has been cited over 40,000 times and is used by most large-scale experimental collaborations\cite{Bierlich:2026syi}.

One of the challenges behind such a large user base is to ensure not only that the code base does not fragment, but also to develop the code to meet diverse user needs.
In some cases, this necessitates the inclusion of external software dependencies and in other cases the inclusion of user-written code that is not core \pythia but still packaged with releases.
In the current \cpp \pythia 8, this has been traditionally accomplished via users including selected header files packaged with \pythia, and then linking against any necessary shared libraries at build time.
In this paper we introduce \contrib, a platform which enables external third-party developers to write code that can then be loaded as a run-time plugin within \pythia; this platform is in part inspired by \fastjet-\textsc{Contrib}\cite{Cacciari:2011ma} hosted at \href{https://www.fastjet.fr/contrib}{\vb{fastjet.fr/contrib}}.
For the purposes of this paper, we define a plugin as a class, derived from a \pythia class, that is not part of the core \pythia library but can be inserted into the \pythia event-generation process at run time.
Plugins are compiled into plugin libraries which are shared libraries, each containing one or more plugins, that \pythia loads at run time.
A package is a collection of one or more plugins distributed through \contrib and corresponds to a repository containing code that builds plugin libraries, together with any associated settings, data, documentation, and examples.

\contrib is intended for three groups: \pythia users, who install packages and use plugins from these packages; package authors, who are developers external to \pythia that write and maintain packages; and internal \pythia authors, who develop the core \pythia code.
\contrib itself is coordinated by the \contrib maintainers, who develop the \contrib system, review new packages for inclusion, and provide minimal updates to packages as needed.
The workflow is as follows:
\begin{enumerate}
\tightlist
\item With the tools documented in \cref{sec:develop}, a package author creates a standalone package, whose plugins are classes derived from \pythia classes.
  This package can use external software dependencies as necessary.
  The package may then be hosted on the \contrib repository.
\item Using the tools described in \cref{sec:use} a user enables the packages providing the plugins they require.
  Packages are built into plugin libraries, independent of the \pythia library.
\item At run time, the user can then load the required plugins from these plugin libraries through \pythia.
\end{enumerate}
In this way, external code and dependencies can be introduced without modifying core \pythia code.
Package authors maintain control over their own code without fragmenting the \pythia code base.
\contrib also provides a clean way for package authors to maintain academic ownership of their implemented model, for example ensuring that their works are properly cited.

\contrib consists of three parts: tools to build and install packages documented in \cref{sec:use}; the run-time plugin system implemented in \pythia and described in \cref{sec:plugins}; and a development kit for package authors detailed in \cref{sec:develop}.
All the code described in this paper is available at \href{https://gitlab.com/pythia8-contrib}{\vb{gitlab.com/pythia8-contrib}} and \href{https://gitlab.com/pythia8/releases}{\vb{gitlab.com/pythia8/releases}}.
The \contrib system is documented on \href{https://contrib.pythia.org}{\vb{contrib.pythia.org}}.
At the time of writing, the current \pythia version was \vb{8.318}.

\section{Installation and Usage of Packages}
\label{sec:use}

A live version of this documentation is available through the \contrib repository at \href{https://gitlab.com/pythia8-contrib/use}{\vb{gitlab.com/pythia8-contrib/use}}.
\input{use}

\section{Technical Implementation of Plugins}
\label{sec:plugins}

\input{plugins}

\section{Development of Packages}
\label{sec:develop}

A live version of the documentation provided here is available through the \contrib repository at \href{https://gitlab.com/pythia8-contrib/develop}{\vb{gitlab.com/pythia8-contrib/develop}} with a package index available at \href{https://contrib.pythia.org}{\vb{contrib.pythia.org}}.
\input{develop}

\section{Conclusions}

In this paper, we introduced \contrib, a run-time plugin system for \pythia that allows external developers to write and distribute code as package authors, without modifying core \pythia or introducing new dependencies.
\contrib spans the full plugin workflow, from development and packaging to run-time loading. 
This keeps the \pythia code base from fragmenting, while letting package authors maintain control over, and receive credit for, their own contributions.
We expect \contrib to grow alongside the \pythia community, and encourage new package authors to contribute plugins through the \contrib repository.

\paragraph*{Acknowledgments.}
We thank the \pythia collaboration for their support of \mlhad work.
We also thank Gavin Salam and the \textsc{PanScales} collaboration for their beta testing of \contrib.
BA, JZ, and RG acknowledge support in part by the DOE grants DE-SC0011784 and DE-SC0026301, and NSF  grants OAC-2103889, OAC-2411215, and OAC-2417682.
JZ, TM, and MW also acknowledge support in part from the Visiting Scholars Award Program of the Universities Research Association.
PI and MW are supported by NSF grants OAC-2103889, OAC-2411215, OAC-2417682, and NSF-PHY-2209769.
CB acknowledges support from Vetenskapsrådet contract number 2023-04316.
Research at the Perimeter Institute is supported by the Government of
Canada through the Department of Innovation, Science and Industry Canada, and by the
Province of Ontario through the Ministry of Colleges and Universities. 
B.A.\ was supported in part by the Canadian AI Safety Institute Research Program at CIFAR. 
This document was prepared using the resources of the Fermi National Accelerator Laboratory (Fermilab), a U.S. Department of Energy, Office of Science, Office of High Energy Physics HEP User Facility.
Fermilab is managed by FermiForward Discovery Group, LLC, acting under Contract No. 89243024CSC000002.

\bibliography{main.bib}

\end{document}

%% file: use.tex
While \contrib packages can be installed separately, per
individual package documentation, this repository provides a unified set
of tools to install \contrib packages. A simple workflow for
installing packages is as follows.

\begin{enumerate}
\def\labelenumi{\arabic{enumi}.}
\item
  Install \pythia 8, if not already installed, and ensure that the
  command \vb{pythia8-config} is available.
\item
  Check out the \vb{use} repository.

\begin{lstlisting}[language=bash,style=shell]
git clone git@gitlab.com:pythia8-contrib/use.git
cd use
\end{lstlisting}
\item
  Use the \vb{enable} script to enable the
  requested packages.

\begin{lstlisting}[language=bash,style=shell]
./enable PACKAGE[/VERSION] ... [OPTIONS]
\end{lstlisting}
\item
  Run the \vb{configure} script to configure all
  the enabled packages.

\begin{lstlisting}[language=bash,style=shell]
./configure [OPTIONS]
\end{lstlisting}
\item
  Build and install the packages to a location known by \pythia at run
  time. This location can be configured for \pythia by setting the
  environment variable \vb{PYTHIA8CONTRIB}.

\begin{lstlisting}[language=bash,style=shell]
make
make install
\end{lstlisting}
\end{enumerate}

Details on steps 3--5 are given in the dedicated sections of this
documentation. Packages can also be configured individually. To do this,
use the same workflow as above, but, in step 4, first configure each
package by hand and then run \vb{configure} at the
top level. Any packages that have already been configured will not be
configured when running the top level
\vb{configure}.

As an example, consider an installation of the
\vb{SimpleCluster} package, which provides a
hadronization plugin using a simplified cluster model. To begin, \pythia
8 must already be installed.

\begin{lstlisting}[language=bash,style=shell]
PYTHIA=<installation path for Pythia 8>
git clone git@gitlab.com:Pythia8/releases.git pythia83
cd pythia83
./configure --prefix=$PYTHIA
make
make install
\end{lstlisting}

Next, \vb{pythia8-config} must be available on
\vb{PATH}. If it is not, update the
\vb{PATH} environment variable.

\begin{lstlisting}[language=bash,style=shell]
export PATH="$PATH:$PYTHIA/bin"
\end{lstlisting}

Note, it is also possible to use the options starting with
\vb{--with-pythia8} as part of the
\vb{configure} script to set the installed location
of \pythia 8.

Now that \pythia 8 is available, we can enable, configure, and build the
\vb{SimpleCluster} package.

\begin{lstlisting}[language=bash,style=shell]
git clone git@gitlab.com:pythia8-contrib/use.git
cd use
./enable SimpleCluster
./configure
make
make install
\end{lstlisting}

Here, we install the enabled \contrib packages along the same
\vb{$PYTHIA} prefix as the \pythia 8 installation,
the default behavior when \vb{pythia8-config} is
available.

Finally, we can run one of the examples provided by the
\vb{SimpleCluster} package. For most packages,
examples should be provided by the package authors and will be copied
into the \vb{share/Pythia8Contrib} directory of the
\pythia 8 installation when \vb{make install} is
called. Here, the examples will have been copied to the directory
\vb{share/Pythia8Contrib/SimpleCluster/examples} in
the \pythia 8 installation.

\begin{lstlisting}[language=bash,style=shell]
export PYTHIA8CONTRIB=$(pythia8-config --datadir)/../Pythia8Contrib
cd $PYTHIA8CONTRIB/SimpleCluster/examples
make testPlugins
./testPlugins testPlugins.cmnd UserFrag.cmnd
\end{lstlisting}

Here, \vb{testPlugins} is an executable where files
containing \pythia settings can be passed as arguments.

\hypertarget{enabling-packages}{%
\subsection{Enabling Packages}\label{enabling-packages}}

Enabling a package checks out the package source code from
\href{https://gitlab.com/pythia8-contrib/packages}{\vb{gitlab.com/pythia8-contrib/packages}}
so that it can be configured, built, and installed. The usage of
\vb{enable} is as follows:

\begin{lstlisting}[language=bash,style=shell]
./enable PACKAGE[/VERSION] ... [OPTIONS]
\end{lstlisting}

A help message can be printed by passing the option
\vb{--help}. Any number of packages with name
\vb{PACKAGE} can be enabled, with the optional
\vb{/VERSION} passed after each
\vb{PACKAGE} to specify the package version. If
\vb{VERSION} for a package is not specified, then
the largest tagged version number for that package will be used. For
example,

\begin{lstlisting}[language=bash,style=shell]
./enable SimpleCluster/v8317.01 Hdsense
\end{lstlisting}

will enable the \vb{SimpleCluster} package, using
\vb{v8317.01} if this version is available, and
enable the \vb{Hdsense} package with the largest
tagged version. If a requested version is not available, then the
largest tagged version number of the package will be used. If there are
no tagged version numbers for the package, then the head of the package
is used. The head of the package, tagged or otherwise, can also be
requested explicitly by setting \vb{VERSION} to
\vb{HEAD}.

The \vb{enable} script can be run multiple times.
However, if a package has already been enabled, then
\vb{enable} will skip that package, unless the
\vb{--force} option is used. In that case,
\vb{enable} deletes the enabled package and checks
out a new copy of the package.

Passing the \vb{--list} option to
\vb{enable} will print all available packages and
disregard all other arguments. An example of the output (at the time of
writing) is given below.

\begin{lstlisting}
The following packages are available from pythia8-contrib.
SherpaShower
Hdsense
IRCExplorer
SherpaFrag
SimpleCluster
ExampleML
\end{lstlisting}

The \vb{--all} option will enable all the packages
listed by \vb{--list}, using the most recent
version available for each package.

The available versions for a package can be listed by passing the
\vb{--tags} option to
\vb{enable}. The tagged versions will then be
printed for each requested \vb{PACKAGE[/VERSION]},
without enabling any of them. For example the command,

\begin{lstlisting}[language=bash,style=shell]
./enable SimpleCluster Hdsense --tags
\end{lstlisting}

will produce the following example output (at the time of writing).

\begin{lstlisting}
SimpleCluster:
    v8317.00
    v8312.00
Hdsense:
    v8316.00
\end{lstlisting}

Here, \vb{SimpleCluster} has two tagged versions,
while \vb{Hdsense} has one tagged version. The
version tagging depends upon the convention adopted by the package
author, but it is suggested that the major version, \emph{e.g.},
\vb{v8317}, specifies the \pythia version the
package was developed for, while the minor version, \emph{e.g.},
\vb{00}, specifies the package version. The tag
does not generally specify compatibility with \pythia versions. Instead,
\pythia version compatibility can be checked either at
\href{https://contrib.pythia.org}{\vb{contrib.pythia.org}}
or through the \vb{configure} script.

\hypertarget{configuring-packages}{%
\subsection{Configuring Packages with External
Dependencies}\label{configuring-packages}}

Packages may have external dependencies that must be configured before
building. \contrib does not require a specific build configuration
system be used for every package, and leaves this as a choice to the
package authors. However, a default build configuration system is
provided. The \vb{configure} script uses this
common system to allow a user to configure the dependencies for packages
with a single command. For packages that do not use this system, please
refer to the documentation for that package and configure it directly.

The \vb{configure} command has the following usage.

\begin{lstlisting}[language=bash,style=shell]
./configure [OPTIONS]
\end{lstlisting}

The \vb{--help} option prints the documentation for
\vb{configure}. Running
\vb{configure} will configure all packages that
have been enabled with the \vb{enable} command and
that are not already configured. A package is marked as configured if
the file \vb{Makefile.inc} is present at the top
level of the package, \emph{i.e.},
\vb{PACKAGE/Makefile.inc}. If the package is
already configured, \vb{configure} will skip
configuring that package and print a warning. Passing the
\vb{--force} option will configure all packages,
even if already configured. The \vb{--versions}
option will print the compatible \pythia 8 versions for each package,
without configuring the package.

By convention, options starting with
\vb{--with-DEP} allow the dependency
\vb{DEP} to be configured, where
\vb{DEP} is all lowercase. For example, the options
starting with \vb{--with-fastjet3} configure the
\fastjet package. The following \vb{--with-DEP}
options allow for fine-grained control over finding the relevant
dependency.

\begin{itemize}
\tightlist
\item
  \vb{--with-DEP=DIR}: look for
  \vb{DEP} at the top directory
  \vb{DIR}.
\item
  \vb{--with-DEP-bin=DIR}: look for
  \vb{DEP}'s binaries along the path
  \vb{DIR}.
\item
  \vb{--with-DEP-lib=DIR}: look for
  \vb{DEP}'s libraries along the path
  \vb{DIR}.
\item
  \vb{--with-DEP-include=DIR}: look for
  \vb{DEP}'s headers along the path
  \vb{DIR}.
\item
  \vb{--with-DEP-config=BIN}: use
  \vb{BIN} as the configuration script for
  \vb{DEP}.
\end{itemize}

Note that the base \pythia build will have its own configuration. Any
packages configured there have no impact on a \contrib package.

An example of enabling external dependencies is as follows.

\begin{lstlisting}[language=bash,style=shell]
./enable Hdsense
./configure --with-fastjet3=<path to FastJet3>
\end{lstlisting}

No checks are made by \vb{configure} to ensure that
the dependency options passed this way are valid. Additionally, by
convention the package-level configurations will also not issue warnings
for invalid options. This is because one package might require a \fastjet
dependency, while another may require HepMC but not \fastjet.

Options other than those beginning with
\vb{--with-DEP} can also be passed to
\vb{configure}. In this way, \cmake or other
build-system options can be passed, like
\vb{-DBUILD\_SHARED\_LIBS=ON}, for example.
\textbf{By necessity, all options are passed to the configuration system
for each enabled package.} This means that
\vb{configure} does not allow different versions of
a dependency for different packages. For example, to configure one
package using \fastjet 3.5.0 and all other packages with \fastjet 3.5.1,
configure the first package manually to use \fastjet 3.5.0 and then run
\vb{configure}, passing \fastjet 3.5.1 as a
dependency, to configure all the other packages.

All packages can be installed after being built. The
\vb{--prefix} and
\vb{--prefix-COMPONENT} options specify where the
packages are installed. If no \vb{--prefix} options
are passed, \vb{configure} will use
\vb{pythia8-config}, if available, to determine the
installation paths. If \vb{pythia8-config} is not
available, \vb{--prefix} defaults to the current
working directory. The \vb{pythia8-config} script
can be made available by placing it on \vb{PATH}.

\begin{lstlisting}[language=bash,style=shell]
export PATH="$PATH:<directory containing pythia8-config>"
\end{lstlisting}

If \vb{pythia8-config} is available,
\vb{configure} will also check if the package is
compatible with the version of \pythia 8 obtained by the command
\vb{pythia8-config --version}. If the package is
not compatible, a warning will be printed.

It is also possible to directly specify the installation path using
\vb{--prefix=DIR} which places all \contrib
components along the common path \vb{DIR}.
Fine-grained installation control is also available with the
\vb{--prefix-bin},
\vb{--prefix-lib},
\vb{--prefix-include}, and
\vb{--prefix-share} options. For example,

\begin{lstlisting}[language=bash,style=shell]
./configure --prefix-share=SHAREDIR
\end{lstlisting}

will install the shared data for each package to the directory
\vb{SHAREDIR} but all other components to the paths
determined from \vb{pythia8-config}. Note that for
the plugins in most packages to be usable by \pythia 8, only the plugin
library and share data need to be installed.

\hypertarget{building-packages}{%
\subsection{Building Packages}\label{building-packages}}

To build all the packages, \vb{make} should be
called after calling \vb{configure}, as done in the
initial workflow above. Any standard options can be passed to make, for
example use all available cores to build.

\begin{lstlisting}[language=bash,style=shell]
make -j
\end{lstlisting}

Before calling \vb{make},
\vb{configure} should be run and a
\vb{Makefile.inc} should be present in the
top-level directory. This \vb{Makefile.inc} is read
by the top-level \vb{Makefile} and sets the
packages to build. If a package has not been successfully configured,
then that package will be skipped. The top-level
\vb{Makefile} calls \vb{make}
in each enabled and configured package with the argument
\vb{contrib} and also passes all options given by
the user to \vb{make} at the top level.

There are a number of targets that can be passed to the
\vb{make} command at the top level. Individual
packages can be built by calling \vb{make PACKAGE},
again with whatever additional options are needed. Calling
\vb{make clean} removes the local build directories
(\vb{bin}, \vb{include},
\vb{lib}, and \vb{share}) but
does not call clean on each package. Calling
\vb{make cleanpkgs} will run
\vb{make clean} for each package while
\vb{make distclean} will remove all
\vb{enable}- and
\vb{configure}-generated files, including the local
build directories and every successfully configured package directory.
Calling \vb{make install} installs the contents of
the local build directories to the locations specified with the
\vb{--prefix} options passed to
\vb{configure}.

\hypertarget{xml-paths}{%
\subsection{Specifying XML Paths for Package Settings}\label{xml-paths}}

Some \contrib packages use the \pythia 8 XML framework to specify
settings and provide documentation. If this is the case, the package
will typically look for the file
\vb{PACKAGE/xmldoc/Index.xml} along the following
paths.

\begin{enumerate}
\def\labelenumi{\arabic{enumi}.}
\tightlist
\item
  Relative to the \vb{PYTHIA8CONTRIB} environment
  variable, \emph{i.e.},
  \vb{$PYTHIA8CONTRIB/PACKAGE/xmldoc/Index.xml}.
\item
  Relative to the \vb{PYTHIA8DATA} environment
  variable, \emph{i.e.},
  \vb{$PYTHIA8DATA/../../PACKAGE/xmldoc/Index.xml}.
\item
  The path itself (whether relative or absolute), \emph{i.e.},
  \vb{PACKAGE/xmldoc/Index.xml}.
\end{enumerate}

If the relevant XML index for the package cannot be found, \pythia 8 will
issue the following warning message.

\begin{lstlisting}
Settings::init: settings file PACKAGE/xmldoc/Index.xml not found
\end{lstlisting}

If this is the case, then set \vb{PYTHIA8CONTRIB}
such that
\vb{$PYTHIA8CONTRIB/PACKAGE/xmldoc/Index.xml}
points to the XML index for that package.

\begin{lstlisting}[language=bash,style=shell]
export PYTHIA8CONTRIB=<path to share/Pythia8Contrib>
\end{lstlisting}

In many cases, \pythia 8 may be installed via a package manager
(\vb{dnf}, \vb{apt},
\vb{PyPI}, \vb{Anaconda},
\emph{etc.}). In these cases the environment variable
\vb{PYTHIA8DATA} might be set by the \pythia 8 setup
from the package manager and may override the XML path for local
versions of \pythia 8, resulting in incompatible versions of the XML
framework being loaded. If this occurs, the
\vb{PYTHIA8DATA} environment variable needs to be
set explicitly to reference the folder containing the top-level
\vb{Index.xml} needed by \pythia.

%% file: plugins.tex
Before describing the plugin system, we first introduce the \pythia user interface and provide a concrete example of the difference between using header and plugin user code.
The user interface to \pythia is primarily through the \vb{Pythia} and \vb{Event} classes.
The following short example demonstrates how events can be generated and analyzed by a user.
\code[language=C++]{figs/ui.cc}
The \vb{Pythia} object can be configured via \vb{Pythia::readString} through command strings like \vb{"SoftQCD:all = on"}, which turns on all soft QCD production processes.
Command strings can also be collected into command files, where each line corresponds to a configuration command, and then read with the \vb{Pythia::readFile} method.
Once the \vb{Pythia} object is configured it must be initialized with \vb{Pythia::init}.
Then, events can be generated via \vb{Pythia::next} and analyzed through the event record, accessible via the \vb{Pythia::event} object.
This \vb{Event} object contains \vb{Particle} objects which correspond to the particles generated by \pythia for that event.

For the analysis of the event record one can use, for instance, the \rivet package, Robust Independent Validation of Experiment and Theory \cite{Bierlich:2024vqo}, which provides a unified method for applying user analyses to the outputs of different MCEGs.
For the example above, this would correspond to generating histograms using the information from \vb{Pythia::event} after each \vb{Pythia::next} call.
In \pythia \vb{8.315} and earlier, the \vb{Pythia8Plugins/Pythia8Rivet.h} header provided the \vb{Pythia8Rivet} helper class for running \rivet analyses on \pythia output.

The following code on the left demonstrates how a \rivet analysis would be run using the header interface with \pythia \vb{8.315} and prior, while the code on the right demonstrates how a \rivet analysis can be run using the plugin interface with \pythia \vb{8.316} and beyond.
\begin{paracol}{2}
  \catcode`\|=0 \catcode`\\=12
  |lstset{escapeinside={\\!}{!}}
  |catcode`\=0 \catcode`\|=12
  \code[language=C++]{figs/rivet-header.cc}
  \switchcolumn
  \code[language=C++]{figs/rivet-plugin.cc}
  \lstset{escapeinside={}{}}
\end{paracol}
\noindent
Building the header example requires linking against all the relevant \rivet libraries and the \pythia library, while building the plugin example requires linking only against the \pythia library, with the \rivet plugin library \vb{libpythia8rivet.so} built independently.
The plugin example automatically handles parallel running when the \vb{Pythia} object is replaced with a \vb{PythiaParallel} object, while the header example does not.

The plugin functionality is available for a number of \pythia base classes which can be passed as external pointers to a \vb{Pythia} object.
At the time of writing, these base classes, the \vb{Pythia} methods that set them, and their documentation are listed in \cref{tab:classes}.

In the remainder of this section more details are given on how users can load plugins (\cref{sec:plugins:load}) and on the technical implementation behind plugins (\cref{sec:plugins:implement}).
\begin{table}[!t]
  \caption{Base classes that can be loaded as plugins, the method of the \pythia class that sets each one, links to the HTML manual and \doxygen documentation for \pythia 8.318, and a short description. The same list is printed by \texttt{./generate -{}-list}, see \cref{sec:develop}.}
  \label{tab:classes}
  \footnotesize
  \input{classes}
\end{table}

\subsection{Loading Plugins}
\label{sec:plugins:load}

There are two ways to load plugins within \pythia code.
The first is to provide a list of plugins to \pythia via the \vb{Init:plugins} setting, as used in the \rivet plugin example above.
As another example, consider writing a \vb{UserHooks} class called \vb{MyUserHooks} and compiling this in the plugin library \vb{libMyPlugins.so}.
In \pythia, the \vb{UserHooks} class provides a way for users to inject custom code into the \pythia event generation process at specific points.
For example, a \vb{UserHooks} class could be used to veto events after the parton shower or modify the cross section of the hard process.
To automatically load this user hook and pass it to a \vb{Pythia} object, the following command should be given to \pythia

\begin{lstlisting}
Init:plugins = {libMyPlugins.so::MyUserHooks, ...}
\end{lstlisting}

where more than one plugin can be specified.
Then, when \vb{Pythia::init()} is called, a \vb{MyUserHooks} object will be automatically created and set as the \vb{UserHooks} object for the \vb{Pythia} object via the method \vb{Pythia::setUserHooksPtr}.
For most plugin classes of generic type \vb{BaseClass}, the relevant pointer will be set via the corresponding \vb{Pythia::setBaseClassPtr} method.
For some base classes, there are options to either set the pointer, overwriting any previous pointer, or add or insert the pointer to a list of pointers that will all be used.
To specify whether the plugin should be set, inserted, or added, a third optional keyword can be specified:

\begin{lstlisting}
Init:plugins = {libMyPlugins.so::MyUserHooks::set}
\end{lstlisting}

where this third keyword can be either \vb{set} (the default), \vb{add}, or \vb{insert(i)}.
Here, \vb{i} specifies the index in the list of pointers where the new pointer should be inserted.
In this example, if there was a previous \vb{UserHooks} pointer, it would be replaced by a new \vb{MyUserHooks} pointer.

The treatment of this third keyword is different for \vb{PDF} plugins, where instead it specifies the beam type since the \vb{PDF} pointer is always set, rather than added. For example,

\begin{lstlisting}
Init:plugins = {libMyPlugins.so::MyPDF1::A, libMyPlugins.so::MyPDF2::B}
\end{lstlisting}

sets \vb{MyPDF1} for beam \vb{A}, while \vb{MyPDF2} is set for beam \vb{B}.
This beam type can be taken from the following list of options, which correspond to the named arguments of the \vb{Pythia::setPDFPtr} method: \vb{A}, \vb{B}, \vb{HardA}, \vb{HardB}, \vb{PomA}, \vb{PomB}, \vb{GamA}, \vb{GamB}, \vb{HardGamA}, \vb{HardGamB}, \vb{UnresA}, \vb{UnresB}, \vb{UnresGamA}, \vb{UnresGamB}, \vb{VMDA}, and \vb{VMDB}.
If omitted, the PDF is set for both \vb{A} and \vb{B}.

If there are new settings which are defined in the \vb{libMyPlugins.so} library, these settings are registered before the \vb{MyUserHooks} object is created.
In this way, it is possible to pass settings to the \vb{MyUserHooks} object from a command file.
This can be done as,

\begin{lstlisting}
Init:plugins = {libMyPlugins.so::MyUserHooks::set::plugin.cmnd}
\end{lstlisting}

where \vb{plugin.cmnd} is a standard command file for \pythia. It is also possible to use the subrun feature of the \pythia command files.

\begin{lstlisting}
Init:plugins = {libMyPlugins.so::MyUserHooks::set::plugin.cmnd::2}
\end{lstlisting}

This command tells \pythia to load the specific commands for subrun \vb{2} from the \vb{plugin.cmnd} file.

Plugin classes of the base class \vb{PhaseSpace} are treated slightly differently than the other base classes as there are no direct set, add, or insert methods for the \vb{Pythia} class.
Instead, the phase-space object is optionally passed through the second argument to the relevant methods for the \vb{SigmaProcess} base class.

\begin{lstlisting}[language={C++}]
bool Pythia::setSigmaPtr(SigmaProcessPtr sigmaPtrIn, PhaseSpacePtr phaseSpacePtrIn = nullptr)
\end{lstlisting}

This is handled through the \vb{Init:plugins} command by first adding a \vb{SigmaProcess} plugin class, directly followed by a \vb{PhaseSpace} plugin class.

\begin{lstlisting}
Init:plugins = {libMyPlugins.so::MySigmaProcess,libMyPlugins.so::MyPhaseSpace}
\end{lstlisting}

In this example, third and fourth keywords can still be specified as described above.

The second method for loading a plugin can be used directly in user code, with or without a \vb{Pythia} instance, using the \vb{make_plugin} function.
Before using the \vb{make_plugin} function, the header \vb{Pythia8/Plugins.h} must be included.
This function mirrors the standard \cpp \vb{make_shared} function and creates a new instance of a shared pointer from a plugin library.

\begin{lstlisting}[language={C++}]
shared_ptr<BaseClass> make_plugin<BaseClass>(string libName, string className, Pythia* pythiaPtr = nullptr, Settings* settingsPtr = nullptr, Logger* loggerPtr = nullptr)
\end{lstlisting}

Here, a shared pointer is created for the \vb{BaseClass} object from the plugin library \vb{libName} using the plugin class \vb{className}.
The final three optional arguments of this function are passed to the constructor for \vb{className}: \vb{pythiaPtr} a full \vb{Pythia} object; \vb{settingsPtr} a \vb{Settings} database object; and \vb{loggerPtr} a \vb{Logger} logging object.
If either \vb{settingsPtr} or \vb{loggerPtr} are null and \vb{pythiaPtr} is not, then the relevant pointer from \vb{pythiaPtr} is used instead.
Plugin classes are always required to take a \vb{Pythia}, \vb{Settings}, and \vb{Logger} pointer in their constructors.
However, none of these pointers is required to be valid.
If a plugin class requires one of these pointers to be valid, it is possible for the author of the plugin library to specify this when creating the plugin class.
If \vb{make_plugin} is called for a plugin class that requires one of these pointers but the passed pointer is invalid, \vb{make_plugin} will return a \vb{nullptr} and report an error if logging is available.

A plugin library may optionally register new settings with the \vb{Settings} object that is later passed to the constructors of its plugin classes.
This allows for a wide range of options to be passed to a plugin object through the \vb{Settings} object as follows.
First, the plugin library settings are registered with a \vb{Settings} object.
Then, the new settings from the plugin library in this \vb{Settings} object can be modified by the user.
Finally, the plugin object is created, with the \vb{Settings} object passed to the constructor so that it can access these specialized settings.

There are two overloaded \vb{make_plugin} methods which allow the user to pass settings to the plugin object.
The first method takes a vector of command strings to pass to the plugin, \vb{cmnds}.

\begin{lstlisting}[language={C++}]
shared_ptr<BaseClass> make_plugin<BaseClass>(string libName, string className, Pythia* pythiaPtr, const vector<string>& cmnds)
\end{lstlisting}

The second method takes a file of commands instead, and optionally a subrun.

\begin{lstlisting}[language={C++}]
shared_ptr<BaseClass> make_plugin<BaseClass>(string libName, string className, Pythia* pythiaPtr, string fileName, int subrun = SUBRUNDEFAULT)
\end{lstlisting}

\subsection{Implementation}
\label{sec:plugins:implement}
\lstset{language={C++}}

The \vb{make_plugin} function is the core routine behind plugin loading and uses a number of functions that are exposed for \texttt{C} linkage so they are accessible for run-time loading via \vb{dlsym}.
The functions for each plugin class are as follows, where \vb{CLASS} is the name of that class.
Each of these functions must be defined once per class.
\begin{itemize}
  \tightlist
\item \vb{bool REQUIRE_PYTHIA_CLASS()}: returns \vb{true} if this plugin class requires a valid \vb{Pythia} pointer for its constructor.
\item \vb{bool REQUIRE_SETTINGS_CLASS()}: returns \vb{true} if this plugin class requires a valid \vb{Settings} pointer for its constructor.
\item \vb{bool REQUIRE_LOGGER_CLASS()}: returns \vb{true} if this plugin class requires a valid \vb{Logger} pointer for its constructor.
\item \vb{const char* TYPE_CLASS()}: returns the name of the \pythia base class for this plugin class.
\item \vb{CLASS* NEW_CLASS(Pythia* pythiaPtr, Settings* settingsPtr, Logger* loggerPtr)}: calls the constructor for this plugin class using \vb{new}.
\item \vb{void DELETE_CLASS(CLASS* ptr)}: calls the destructor for this plugin class using \vb{delete}.
\end{itemize}
There are also functions which must be defined only once per plugin library.
\begin{itemize}
  \tightlist
\item \vb{bool CHECK_COMPATIBLE_VERSION(int ver)}: returns \vb{true} if \vb{ver} is a compatible \pythia version with this plugin library.
\item \vb{bool CHECK_COMPILED_VERSION(int ver)}: returns \vb{true} if \vb{ver} is the same \pythia version that this plugin library was compiled against.
\end{itemize}
Finally, there are optional functions that if defined, should only be defined once per plugin library.
\begin{itemize}
  \tightlist
\item \vb{bool CHECK_PARALLEL()}: if defined and returns \vb{true}, then this plugin library can be used in parallel running with \vb{PythiaParallel}.
\item \vb{void REGISTER_SETTINGS(Settings* settingsPtr)}: registers any settings for the library with the passed \vb{Settings} object.
\item \vb{const char* RETURN_XML()}: returns the path to the index file for optional \xml documentation.
\end{itemize}

Loading and calling these functions from the plugin library, the \vb{make_plugin} function then works as follows.
\begin{enumerate}
  \tightlist
\item Check if the passed \vb{settingsPtr} and \vb{loggerPtr} arguments are valid. If not and \vb{pythiaPtr} is valid, then use \vb{pythiaPtr->settings} or \vb{pythiaPtr->logger} respectively.
\item Load the plugin library with \vb{dlopen_plugin}.

\begin{lstlisting}
shared_ptr<void> dlopen_plugin(string libName, Logger* loggerPtr)
\end{lstlisting}

  \begin{enumerate}
  \item Load the plugin library with the \posix function \vb{dlopen} using \vb{RTLD_LAZY}.
  \item If not successfully loaded, report an error and return a null pointer.
  \item Load the \vb{CHECK_COMPATIBLE_VERSION} symbol from the plugin library using the \posix function \vb{dlsym}.
    If not available, report an error and return a null pointer.
  \item Use \vb{CHECK_COMPATIBLE_VERSION} to check the loading version of \pythia is compatible with the plugin library.
    If not, report an error and return a null pointer. 
  \item Load and call \vb{CHECK_COMPILED_VERSION} to check if the \pythia version the plugin library was compiled with matches the \pythia version loading the plugin library.
    If not, report a warning as there may be ABI compatibility issues.
  \item Return a shared pointer to this plugin library with a custom destructor which calls \vb{dlclose} and unloads the plugin library.
  \end{enumerate}
\item Check the base class of the plugin class matches the requested \vb{BaseClass} by loading and calling \vb{TYPE_CLASS}.
  If not, report an error and return a null pointer.
\item Load and call \vb{REQUIRE_PTR_CLASS}, where \vb{PTR} is \vb{PYTHIA}, \vb{SETTINGS}, or \vb{LOGGER}.
  If any of these symbols do not exist, or if the function returns \vb{true} and the corresponding \vb{Pythia}, \vb{Settings}, or \vb{Logger} pointer is invalid, then report an error and return a null pointer.
\item Create a shared pointer where the plugin class object is constructed by loading and calling \vb{NEW_CLASS}, and a custom destructor for this shared pointer is defined which loads and calls \vb{DELETE_CLASS}.
\end{enumerate}

When using the \vb{Init:plugins} command the \vb{Settings} class loads the specified plugin libraries using the following method.

\begin{lstlisting}
bool Settings::registerPluginLibrary(string libName, string startFile)
\end{lstlisting}

Here, \vb{libName} is the name of the plugin library, and \vb{startFile} optionally is the path to the index file for associated \xml documentation.
This method then performs the following operations.
\begin{enumerate}
  \tightlist
\item Check if the plugin library is already loaded in the \vb{pluginLibraries} map.
  If the plugin library is already loaded, return \vb{false}.
\item Load the plugin library with \vb{dlopen_plugin}.
  Return \vb{false} if unsuccessful.
\item Try to load and call \vb{RETURN_XML} and set \vb{startFile} if passed as an empty argument.
\item Load and call \vb{CHECK_PARALLEL} to check if the plugin library is compatible with running in parallel through \vb{PythiaParallel}.
  If parallel running has been enabled by the user but the library is not compatible, then report an error and return \vb{false}.
\item Load settings from the optional \xml for the plugin library.
\item Try to load and call \vb{REGISTER_SETTINGS} to register any defined settings for the plugin library.
\item Return \vb{true}.
\end{enumerate}
In this way the settings for each plugin library are only loaded once and are loaded before any plugin classes are constructed.

Individual plugin classes are constructed and configured in the following \vb{Pythia} class method.

\begin{lstlisting}
void Pythia::initPlugins()
\end{lstlisting}

This method is called within \vb{Pythia::init} and performs the following steps.
\begin{enumerate}
\item Create \vb{pdfs}, a map of \vb{PDF} shared pointers corresponding to the arguments of \vb{Pythia::setPDFPtr}.
\item Loop over the requested plugin classes.
\item Load the plugin library with \vb{dlopen_plugin}, continue if unsuccessful.
\item Check the correct base class for the plugin class with \vb{TYPE_CLASS}.
\item If the base class is of type \vb{PDF} then update the \vb{pdfs} map with \vb{make_plugin<PDF>} and the third keyword for this plugin class, \eg \vb{PomA}.
\item If the base class is any other type, then call the relevant \vb{Pythia::setBaseClassPtr}, \vb{Pythia::addBaseClassPtr}, or \vb{Pythia::insertBaseClassPtr} methods with the argument \vb{make_plugin<BaseClass>}.
\item After the loop is completed, \vb{Pythia::setPDFPtr} is called with arguments from the \vb{pdfs} map.
\end{enumerate}
\lstset{language={}}

%% file: classes.tex
\begin{tabularx}{\linewidth}{@{}lll>{\raggedright\arraybackslash}X@{}}
  \toprule
  base class & \vb{Pythia} method & doc. & description \\
  \midrule
  \vb{BeamShape} & \vb{setBeamShapePtr} & \href{https://pythia.org/manuals/pythia8318/BeamShape.html}{HTML} \href{https://pythia.org/doxygen/pythia8318/classPythia8_1_1BeamShape.html}{\dox} & custom shape for the incoming beams \\[3pt]
  \vb{DecayHandler} & \vb{setDecayPtr} & \href{https://pythia.org/manuals/pythia8318/ExternalDecays.html}{HTML} \href{https://pythia.org/doxygen/pythia8318/classPythia8_1_1DecayHandler.html}{\dox} & decays using an external tool \\[3pt]
  \vb{FragmentationModel} & \vb{setFragmentationPtr}$^\dagger$ & \href{https://pythia.org/manuals/pythia8318/ImplementNewFragmentation.html}{HTML} \href{https://pythia.org/doxygen/pythia8318/classPythia8_1_1FragmentationModel.html}{\dox} & custom fragmentation model to replace the string model \\[3pt]
  \vb{HeavyIons} & \vb{setHeavyIonsPtr} & \href{https://pythia.org/manuals/pythia8318/HeavyIons.html}{HTML} \href{https://pythia.org/doxygen/pythia8318/classPythia8_1_1HeavyIons.html}{\dox} & fully generalized heavy-ion model \\[3pt]
  \vb{HIUserHooks} & \vb{setHIHooks} & \href{https://pythia.org/manuals/pythia8318/HeavyIons.html}{HTML} \href{https://pythia.org/doxygen/pythia8318/classPythia8_1_1HIUserHooks.html}{\dox} & \vb{UserHooks} and \vb{MergingHooks}, but for heavy-ion collisions \\[3pt]
  \vb{LHAup} & \vb{setLHAupPtr} & \href{https://pythia.org/manuals/pythia8318/LHA.html}{HTML} \href{https://pythia.org/doxygen/pythia8318/classPythia8_1_1LHAup.html}{\dox} & Les Houches Accord user-provided process \\[3pt]
  \vb{Merging} & \vb{setMergingPtr} & \href{https://pythia.org/manuals/pythia8318/ImplementNewMatching.html}{HTML} \href{https://pythia.org/doxygen/pythia8318/classPythia8_1_1Merging.html}{\dox} & bridge between the parton-shower history and \pythia \\[3pt]
  \vb{MergingHooks} & \vb{setMergingHooksPtr} & \href{https://pythia.org/manuals/pythia8318/ImplementNewMatching.html}{HTML} \href{https://pythia.org/doxygen/pythia8318/classPythia8_1_1MergingHooks.html}{\dox} & parton-shower matching/merging scheme \\[3pt]
  \vb{PDF} & \vb{setPDFPtr} & \href{https://pythia.org/manuals/pythia8318/PartonDistributions.html}{HTML} \href{https://pythia.org/doxygen/pythia8318/classPythia8_1_1PDF.html}{\dox} & parton distribution functions for the incoming beams \\[3pt]
  \vb{PhaseSpace} & \vb{setSigmaPtr} (arg.~2)$^\dagger$ & \href{https://pythia.org/manuals/pythia8318/SemiInternalProcesses.html}{HTML} \href{https://pythia.org/doxygen/pythia8318/classPythia8_1_1PhaseSpace.html}{\dox} & phase-space integrator for a custom \vb{SigmaProcess}; cannot be used by itself \\[3pt]
  \vb{ResonanceWidths} & \vb{setResonancePtr}$^\dagger$ & \href{https://pythia.org/manuals/pythia8318/SemiInternalResonances.html}{HTML} \href{https://pythia.org/doxygen/pythia8318/classPythia8_1_1ResonanceWidths.html}{\dox} & external width calculator for resonances \\[3pt]
  \vb{RndmEngine} & \vb{setRndmEnginePtr} & \href{https://pythia.org/manuals/pythia8318/RandomNumbers.html}{HTML} \href{https://pythia.org/doxygen/pythia8318/classPythia8_1_1RndmEngine.html}{\dox} & random number generator used by all of \pythia \\[3pt]
  \vb{ShowerModel} & \vb{setShowerModelPtr} & \href{https://pythia.org/manuals/pythia8318/ImplementNewShowers.html\#anchor7}{HTML} \href{https://pythia.org/doxygen/pythia8318/classPythia8_1_1ShowerModel.html}{\dox} & custom parton shower, providing both a timelike (FSR) and a spacelike (ISR) shower \\[3pt]
  \vb{SigmaProcess} & \vb{setSigmaPtr} (arg.~1)$^\dagger$ & \href{https://pythia.org/manuals/pythia8318/SemiInternalProcesses.html}{HTML} \href{https://pythia.org/doxygen/pythia8318/classPythia8_1_1SigmaProcess.html}{\dox} & new user-provided hard process, optionally used in conjunction with \vb{PhaseSpace} \\[3pt]
  \vb{UserHooks} & \vb{setUserHooksPtr}$^\dagger$ & \href{https://pythia.org/manuals/pythia8318/UserHooks.html}{HTML} \href{https://pythia.org/doxygen/pythia8318/classPythia8_1_1UserHooks.html}{\dox} & omnibus class which allows interactions with the \pythia event-generation sequence at a number of defined steps \\
  \bottomrule
\end{tabularx}

\smallskip
\raggedright
$^\dagger$\,The corresponding \vb{add} and \vb{insert} methods, \emph{e.g.}, \vb{addUserHooksPtr}, are also available. \vb{setSigmaPtr} takes the \vb{SigmaProcess} as its first argument and an optional \vb{PhaseSpace} as its second.

%% file: develop.tex
The tools of this
\href{https://gitlab.com/pythia8-contrib/develop}{\vb{develop}}
repository are intended to provide a convenient way to develop packages
for \pythia 8 that can be installed through the
\href{https://gitlab.com/pythia8-contrib/use}{\vb{use}}
repository of \contrib. For more details on how the plugin system
for \pythia 8 works, specifically how it can be configured by the user,
please consult the
\href{https://pythia.org/manuals/pythia8318/Plugins.html}{HTML manual}.
For a new package to be made public via
\href{https://gitlab.com/pythia8-contrib/use}{\vb{use}},
the package must first be submitted, see the
\protect\hyperlink{package-submission}{package submission} section, and
a corresponding repository created in the
\href{https://gitlab.com/pythia8-contrib/packages}{\vb{packages}}
repository, see the \protect\hyperlink{repository-creation}{repository
creation} section. The basic usage of the
\href{https://gitlab.com/pythia8-contrib/develop}{\vb{develop}}
repository is to create a \contrib package template using the
syntax,

\begin{lstlisting}[language=bash,style=shell]
./generate PACKAGE BASE/NAME ...
\end{lstlisting}

which will create a ready-to-compile package template with name
\vb{PACKAGE} and classes of base-class type
\vb{BASE} with corresponding names
\vb{NAME}, given by the list of
\vb{BASE/NAME} pairs following the package name. By
default, the \vb{BASE} arguments should be taken
from the list of available base classes which can be passed to \pythia as
a plugin using the \vb{Init:plugins} setting.

\begin{lstlisting}
Init:plugins = {libPACKAGE.so::NAME}
\end{lstlisting}

This list of base-class templates can be printed with,

\begin{lstlisting}[language=bash,style=shell]
./generate --list
\end{lstlisting}

which reproduces the list of base classes in \cref{tab:classes}, where
links to further documentation for each base class are also given. Of
note, the \vb{UserHooks} class provides an omnibus
interface to most parts of the \pythia event generation sequence, and
should be considered when a more specific base class is not available.

For the eager package author, a very abbreviated set of commands is
given below that will generate, compile, and run an example
\contrib package, assuming that \pythia is already installed and
\vb{pythia8-config} is available as a command.

\begin{lstlisting}[language=bash,style=shell]
# Generate the template for two user classes, MyUserHooks and MyPDF.
./generate MyPackage UserHooks/MyUserHooks PDF/MyPDF
# Build the package.
cd MyPackage
make -j
# Set the PYTHIA8CONTRIB path.
export PYTHIA8CONTRIB=$PWD/share
# Compile and run the test example for MyUserHooks.
cd share/MyPackage/examples/
make testPlugins
./testPlugins MyUserHooks.cmnd
\end{lstlisting}

Further details for each step are given in the remainder of this
documentation.

\hypertarget{package-generation}{%
\subsection{Package Generation}\label{package-generation}}

The output of \vb{generate} will produce a folder
with name \vb{PACKAGE}, with the following
structure. Each item of this list is described in further detail in its
own dedicated section below.

\begin{itemize}
\tightlist
\item
  \protect\hyperlink{headers}{\vb{include/PACKAGE}}:
  this directory is where all headers for the package are kept. For each
  \vb{BASE/NAME} pair there is a header file
  \vb{NAME.h}.
\item
  \protect\hyperlink{source}{\vb{src}}: this
  directory is where all non-header source code is kept. For every
  \vb{BASE/NAME} pair there is a source file
  \vb{NAME.cc}. The additional file
  \protect\hyperlink{plugin-library-definition}{\vb{Package.cc}}
  defines the properties of the plugin library.
\item
  \protect\hyperlink{additional-documentation}{\vb{share/PACKAGE}}:
  this directory can be used for any additional data or documentation.
  For example, with \pythia, this directory holds the XML and HTML
  documentation, as well as examples and other necessary data.
\item
  \protect\hyperlink{build-system}{\vb{configure},
  \vb{Makefile}}: the
  \vb{configure} script and
  \vb{Makefile} specify how the package is built.
\item
  \protect\hyperlink{documentation}{\vb{README.md}}:
  top-level documentation for the package in Markdown format.
\end{itemize}

\hypertarget{example}{%
\subsubsection{Example}\label{example}}

As an initial example, the following command will produce a
\contrib package with the name \vb{MyPackage}
consisting of: a \vb{MyUserHooks} class of
base-class type \vb{UserHooks} and a
\vb{MyPDF} class of base-class type
\vb{PDF}.

\begin{lstlisting}[language=bash,style=shell]
./generate MyPackage UserHooks/MyUserHooks PDF/MyPDF
\end{lstlisting}

The package structure for this example is as follows.

\begin{lstlisting}
MyPackage/
├── configure
├── .configure.inc
├── .gitignore
├── include
│   └── MyPackage
│       ├── MyPDF.h
│       └── MyUserHooks.h
├── Makefile
├── .meta.yaml
├── README.md
├── share
│   └── MyPackage
│       ├── examples
│       │   ├── Makefile
│       │   ├── MyPDF.cmnd
│       │   ├── MyUserHooks.cmnd
│       │   └── testPlugins.cc
│       └── xmldoc
│           ├── Formatting.xml
│           ├── Index.xml
│           └── Settings.xml
└── src
    ├── MyPDF.cc
    ├── MyUserHooks.cc
    └── Package.cc
\end{lstlisting}

By default only base classes from the list given by
\vb{./generate --list} can be provided as
\vb{BASE} arguments to the
\vb{BASE/NAME} list. If a base class is not
available, \vb{generate} will skip that
\vb{BASE/NAME} pair and issue the following
warning.

\begin{lstlisting}[language=bash,style=shell]
./generate PACKAGE BASE/NAME
WARNING: Unknown template class 'BASE', skipping 'BASE/NAME'.
\end{lstlisting}

If the option \vb{--nocheck} is specified, then the
\vb{BASE/NAME} pair will not be skipped and a
minimal template will be produced for that
\vb{BASE/NAME} pair. For more details on working
with classes that cannot be loaded via the
\vb{Init:plugins} setting, please see the
\protect\hyperlink{additional-classes}{additional classes} section.

\hypertarget{template-code}{%
\subsubsection{Template Code}\label{template-code}}

For each base-class type, the default behavior of
\vb{generate} is to produce template code where all
inherited base-class methods are provided. All methods that need to be
defined, \emph{i.e.}, pure virtual methods of the base class, are
implemented with an ansatz that should be changed by the package author.
Methods that are implemented in the base class but may be overridden are
provided as comments. In the example above, the following method of
\vb{MyPDF} must be implemented,

\begin{lstlisting}[language={C++}]
  // Update parton densities (defined in src/MyPDF.cc). Must be defined.
  void xfUpdate(int id, double x, double Q2) override;
\end{lstlisting}

while the method,

\begin{lstlisting}[language={C++}]
  // // Allow for new scaling factor for VMD PDFs.
  // void setVMDscale(double = 1.) override {}
\end{lstlisting}

is optional. The constructor and destructor for every package class must
also be defined,

\begin{lstlisting}[language={C++}]
  // Constructor (defined in src/MyPDF.cc).
  MyPDF(Pythia*, Settings*, Logger*);
  
  // Destructor.
  ~MyPDF() = default;
\end{lstlisting}

where a template signature is provided in the header file and a template
definition in the source file. The templates produced with the default
\vb{generate} behavior compile out-of-the-box.

\hypertarget{minimal-template}{%
\subsubsection{Minimal Template}\label{minimal-template}}

In some cases, the package author might wish to begin with a minimal
template, where only the constructor and destructor are already defined.
This is possible by passing the \vb{--minimal}
option to \vb{generate}. The template code
generated here will not compile out-of-the-box, as the pure virtual
methods of the base class will need to be implemented by the package
author. When using the \vb{--nocheck} option above,
full template code will be produced if the base class is known, and by
necessity, minimal template code will be produced if the base class is
not known.

\hypertarget{modifying-a-package}{%
\subsubsection{Modifying a Package}\label{modifying-a-package}}

Because the \vb{generate} command could potentially
overwrite already written code, the default behavior of
\vb{generate} is to fail when asked to produce the
template for a package directory that already exists. For example,
running the command

\begin{lstlisting}[language=bash,style=shell]
./generate PACKAGE BASE/NAME
\end{lstlisting}

twice will result in the following error message.

\begin{lstlisting}[language=bash,style=shell]
ERROR: The 'PACKAGE' package already exists, use '--update' to update.
\end{lstlisting}

Here, it is possible to update an already generated package with
additional classes using the \vb{--update} option.

\begin{lstlisting}[language=bash,style=shell]
./generate --update PACKAGE BASE/NAME
\end{lstlisting}

Any class of name \vb{NAME} being requested that
already exists in the package will be skipped with the following
warning.

\begin{lstlisting}[language=bash,style=shell]
WARNING: Skipping existing class name 'NAME'.
\end{lstlisting}

\hypertarget{testing-a-package}{%
\subsection{Testing a Package}\label{testing-a-package}}

To compile and test a package, a working installation of \pythia is
required. The following steps are intended to quickly set up such an
environment, and should give a basic outline of the process. For more
details on installing \pythia, please consult the
\href{https://gitlab.com/Pythia8/releases/-/blob/pythia8318/README}{\vb{README}}
included as part of the \pythia source.

\hypertarget{build-pythia}{%
\subsubsection{Build \pythia}\label{build-pythia}}

\pythia can be installed by first checking out the \pythia
\vb{releases} repository.

\begin{lstlisting}[language=bash,style=shell]
git clone git@gitlab.com:Pythia8/releases.git
cd releases
# Optionally check out a specific version.
# git checkout pythia8318
\end{lstlisting}

It is possible to check out a specific version of \pythia, but only
versions \vb{8.310} and higher are compatible with
the plugin structure of \contrib. Next, configure, build, and
install \pythia.

\begin{lstlisting}[language=bash,style=shell]
./configure  # Include optional configuration flags here. 
make -j      # Build with an optional number of cores.
make install # Use --prefix options with configure to change installation.
cd ../
\end{lstlisting}

By default, the \pythia installation will be placed in the directory in
which it was built when \vb{make install} is
called, unless \vb{--prefix} options are passed to
\vb{configure}.

\hypertarget{build-the-package}{%
\subsubsection{Build the Package}\label{build-the-package}}

Once \pythia is installed, make sure that the
\vb{pythia8-config} script can be found by the
\vb{configure} script of the \contrib package
being developed. This can either be done by adjusting the system
\vb{PATH} so that
\vb{pythia8-config} is along it, or by directly
passing the location of the \vb{pythia8-config}
script to \vb{configure}. Assuming the installation
directory of the previous section, the system path can be updated as
follows.

\begin{lstlisting}[language=bash,style=shell]
export PATH=$PATH:$PWD/releases/bin
\end{lstlisting}

Alternatively, the package \vb{configure} script
can be called with the location specified.

\begin{lstlisting}[language=bash,style=shell]
cd MyPackage
./configure --with-pythia8-config=../releases/bin/pythia8-config
\end{lstlisting}

Once the path to \vb{pythia8-config} is specified
using either method, the package can be built with
\vb{make} called from the generated
\vb{MyPackage} directory.

\begin{lstlisting}[language=bash,style=shell]
cd MyPackage # If not already there from calling ./configure.
make -j      # Build with an optional number of cores.
\end{lstlisting}

If the package was built successfully, then the library
\vb{lib/libPACKAGE.so} should exist. In this case,
the command above builds the library
\vb{lib/libMyPackage.so}.

\hypertarget{test-the-package}{%
\subsubsection{Test the Package}\label{test-the-package}}

Minimal tests for the package are included in the template. First, set
the XML path for \contrib, and then change to the examples
directory of the package.

\begin{lstlisting}[language=bash,style=shell]
export PYTHIA8CONTRIB=$PWD/share
cd share/MyPackage/examples
\end{lstlisting}

There is a simple test which can be used to check the loading of each
plugin class and to read in settings, either directly specified by the
plugin library, or via XML documentation provided by the package. The
test takes as arguments command files to pass to \pythia with the
\vb{readFile} command. For each plugin class
provided by the package there is a corresponding command file which
instructs \pythia to load the plugin. In this example, there are two such
configurations, \vb{MyPDF.cmnd} and
\vb{MyUserHooks.cmnd}. First we test the
\vb{PDF} class.

\begin{lstlisting}[language=bash,style=shell]
make testPlugins
./testPlugins MyPDF.cmnd
\end{lstlisting}

The output of this test should be the \pythia welcome banner, as well as
the settings that were introduced by the package, both directly through
the plugin library as well as the XML files. The setting output should
look as follows.

\begin{lstlisting}
 PYTHIA Info from Pythia8Contrib::MyPackage::MyPDF::MyPDF: library setting MyPackage:registered is 1
 PYTHIA Info from Pythia8Contrib::MyPackage::MyPDF::MyPDF: XML setting MyPackage:flag is 0
 PYTHIA Info from Pythia8Contrib::MyPackage::MyPDF::MyPDF: XML setting MyPackage:modeopen is 0
 PYTHIA Info from Pythia8Contrib::MyPackage::MyPDF::MyPDF: XML setting MyPackage:modepick is 0
 PYTHIA Info from Pythia8Contrib::MyPackage::MyPDF::MyPDF: XML setting MyPackage:parm is 0.000000
 PYTHIA Info from Pythia8Contrib::MyPackage::MyPDF::MyPDF: XML setting MyPackage:word is value
 PYTHIA Info from Pythia8Contrib::MyPackage::MyPDF::MyPDF: XML setting MyPackage:fvec is {1,0}
 PYTHIA Info from Pythia8Contrib::MyPackage::MyPDF::MyPDF: XML setting MyPackage:mvec is {0,1}
 PYTHIA Info from Pythia8Contrib::MyPackage::MyPDF::MyPDF: XML setting MyPackage:pvec is {0.000000,1.000000}
 PYTHIA Info from Pythia8Contrib::MyPackage::MyPDF::MyPDF: XML setting MyPackage:wvec is {value1,value2}
\end{lstlisting}

If any issues are encountered, there will be additional output
describing the problem. If the error
\vb{cannot open shared object file} is returned,
please see the section on \protect\hyperlink{library-paths}{library
paths}. Next, we can test the \vb{UserHooks} class
and check that the output is similar.

\begin{lstlisting}[language=bash,style=shell]
./testPlugins MyUserHooks.cmnd
\end{lstlisting}

\hypertarget{documentation}{%
\subsection{Documentation}\label{documentation}}

Documentation for the package should be provided in a top-level
\vb{README.md} as a Markdown file. Additional
documentation can be provided by the package authors using the mechanism
of their choice, but at a minimum, \vb{README.md}
is required. An optional metadata description can also be provided at
the top level of the package with a \vb{.meta.yaml}
YAML file. If this file is provided, it will be parsed when building the
index of all \contrib packages which are then displayed on the
\protect\hyperlink{index-generation}{\vb{PYTHIA-CONTRIB summary pages}}.
The keywords and syntax of this metadata file are:

\begin{itemize}
\item
  \vb{description}: should provide a brief summary
  of the package and its use. More details can be given elsewhere in the
  \vb{README.md}.
\item
  \vb{authors}: provides the authors of the
  package, with \vb{name} required per author.
  Optionally, both an \vb{email} and an
  \vb{orcid} (\href{https://orcid.org/}{ORCID}) can
  also be specified per author. Multiple authors can be included as
  follows, where collaborations can also be added as authors:

\begin{lstlisting}
 authors:
   - name: PACKAGE Collaboration
   - name: First Author
     email: first.author@contrib.pythia.org
     orcid: 0000-0000-0000-0000
   - name: Second Author
     email: second.author@contrib.pythia.org
\end{lstlisting}
\item
  \vb{license}: defines the license for the
  package. It is recommended that a license is defined for every
  package, \emph{e.g.},
  \vb{license: GNU GPL v2 or later}.
\item
  \vb{keywords}: help categorize the use of the
  package. Any number of keywords can be provided. However, it is
  suggested to first check if suitable keywords have already been used
  by other packages for consistency.

\begin{lstlisting}
  keywords:
    - first key word
    - second key word
\end{lstlisting}
\item
  \vb{dependencies}: define the dependencies of the
  package. If a package uses the
  \protect\hyperlink{configuration-customization}{default configuration
  system} then the specifications given in
  \vb{.configure.inc} take precedence. The syntax
  \vb{>=} can be used to specify a version number
  and \vb{(optional)} to indicate that dependency
  is not required.

\begin{lstlisting}
  dependencies:
    - FIRST>=0.0
    - SECOND (optional)
\end{lstlisting}
\item
  \vb{test}: defines a test that can be run during
  indexing of the \contrib packages to check whether the package
  builds and produces expected results. This test can be triggered by
  the \contrib maintainers, and builds the package against the
  most recent internal version of \pythia. This feature is currently
  experimental and is not guaranteed to remain stable. See
  \protect\hyperlink{testing}{package testing} for more details.
  Accepted keys follow the same conventions as GitLab CI and are:
  \vb{image}, \vb{variables},
  \vb{script} (required),
  \vb{timeout}, and
  \vb{allow\_failure}. The
  \vb{image} key is currently displayed on the
  \contrib index pages, if provided.

\begin{lstlisting}
  test:
    image: pythia8/dev:test
    variables:
      NAME: VALUE
    script:
      - COMMAND
    timeout: TIME
\end{lstlisting}
\end{itemize}

\hypertarget{additional-documentation}{%
\subsubsection{Additional
Documentation}\label{additional-documentation}}

In \pythia, documentation and data are provided through the
\vb{share/Pythia8} directory which consists of a
number of sub-directories.

\begin{itemize}
\tightlist
\item
  \vb{htmldoc}: is an interactive HTML manual for
  \pythia, automatically generated from XML documentation.
\item
  \vb{xmldoc}: provides the source for the HTML
  manual. However, it also defines all the settings read in by \pythia at
  construction. The XML documentation is necessary for \pythia to run.
\item
  \vb{pdfdoc}: is a number of PDF documents
  covering more detailed topics and includes the full published physics
  manual.
\item
  \vb{pdfdata}: holds all the data needed for a
  number of parton distribution functions.
\item
  \vb{settings}: consists of command files which
  can be used for \pythia tunes.
\item
  \vb{examples}: contains a comprehensive set of
  example code which covers a wide range of \pythia use cases.
\end{itemize}

In the context of \contrib, it is expected that while package
authors might want to follow a structure similar to \pythia, they may
also have their own preferences on how to provide documentation for
their package. Consequently, other than the required
\protect\hyperlink{documentation}{documentation} described above,
documentation of the package is left to the package authors. However, in
the template package, explicit outlines are given for
\vb{xmldoc} and \vb{examples}
as both of these are considered to be sufficiently important: XML
documentation because it can be used to define settings, and example
programs as these can be used to help test the package as well as
provide concrete scenarios for users.

\hypertarget{xml-settings}{%
\subsubsection{XML Settings}\label{xml-settings}}

Settings for a package may either be included directly in code through
the \vb{PYTHIA8\_PLUGIN\_SETTINGS} macro (see
\protect\hyperlink{package-development}{package development}), or via
XML settings. For the XML settings of a package to be enabled, the
starting XML file must be specified by the line

\begin{lstlisting}[language={C++}]
PYTHIA8_PLUGIN_XML("PACKAGE/xmldoc/Index.xml") 
\end{lstlisting}

which should be included in the package source (typically
\vb{src/Package.cc}). This starting index file is
then searched for along the following paths, in this order:

\begin{enumerate}
\def\labelenumi{\arabic{enumi}.}
\tightlist
\item
  Relative to the \vb{PYTHIA8CONTRIB} environment
  variable, \emph{i.e.},
  \vb{$PYTHIA8CONTRIB/PACKAGE/xmldoc/Index.xml}.
\item
  Relative to the \vb{PYTHIA8DATA} environment
  variable, \emph{i.e.},
  \vb{$PYTHIA8DATA/../../PACKAGE/xmldoc/Index.xml}.
\item
  The path itself (whether relative or absolute), \emph{i.e.},
  \vb{PACKAGE/xmldoc/Index.xml}.
\end{enumerate}

If the macro \vb{PYTHIA8\_PLUGIN\_XML} is used in
the package source but the specified XML index cannot be found, the
following error will be issued when loading the plugin library:

\begin{lstlisting}
PYTHIA Error in Settings::init: settings file PACKAGE/xmldoc/Index.xml not found
\end{lstlisting}

When the index is successfully found, all files specified with the
\vb{<aidx href="FILE">TITLE</aidx>} tags within the
index will be loaded by the \vb{init} method of the
\pythia
\href{https://pythia.org/latest-manual/SettingsScheme.html}{\vb{Settings}}
object. In the template, the following files are included:

\begin{lstlisting}[language=XML]
<aidx href="Formatting">Formatting</aidx><br/>
<aidx href="Settings">Settings</aidx><br/>
\end{lstlisting}

Here, the files \vb{PACKAGE/xmldoc/Formatting.xml}
and \vb{PACKAGE/xmldoc/Settings.xml} will be read
via the \vb{Settings} initialization, and any
settings defined in these XML files will then be registered.

A number of settings types are available through the \pythia XML settings
scheme. It is recommended that all newly defined settings take the
prefix \vb{PACKAGE}, where
\vb{PACKAGE} is the name of the package. This
ensures that there will be no settings collisions with other
\contrib packages, or with \pythia itself. For each setting type, a
number of arguments can be specified. At a minimum, the name of the
setting must be specified with the \vb{name}
argument and the default value with the
\vb{default} argument. For settings which are
numerical, optional \vb{min} and
\vb{max} arguments may be provided to limit the
values the user may specify. If either of these arguments is missing,
then the value will not be limited from below
(\vb{min}) or above
(\vb{max}).

\begin{lstlisting}[language=XML]
<flag name="PACKAGE:flag" default="off">
This is an example boolean setting which can take the values of either "on" or "off".
</flag>
\end{lstlisting}

\begin{lstlisting}[language=XML]
<modeopen name="PACKAGE:modeopen" default="0" min="0" max="10">
This is an example integer setting that can take any integer value between the limits.
</modeopen>
\end{lstlisting}

\begin{lstlisting}[language=XML]
<modepick name="PACKAGE:modepick" default="0" min="0" max="10">
This is also an example integer setting, the same as "modeopen", but now specific values can be defined with the "option" tags below. Note, these options are not technically enforced, i.e. a user can still choose an integer value not specified below, as long as it remains between the limits.
<option value="0">Example defined option.</option>
</modepick>
\end{lstlisting}

\begin{lstlisting}[language=XML]
<parm name="PACKAGE:parm" default="0.0" min="0.0" max="10.0">
This is an example double-precision setting which can take real-number values between the limits.
</parm>
\end{lstlisting}

\begin{lstlisting}[language=XML]
<word name="PACKAGE:word" default="value">
This is an example character-string setting which can take any string value without blanks.
</word>
\end{lstlisting}

Variable-length lists (or vectors) for each of the above setting types,
corresponding to boolean (\vb{flag}), integer
(\vb{mode}), double-precision
(\vb{parm}), and string
(\vb{word}) values, can also be specified using the
\vb{vec} variations below. The general format for
these settings is \vb{\{value1, value2, ...\}}.

\begin{lstlisting}[language=XML]
<fvec name="PACKAGE:fvec" default="{on,off}">
This is an example variable length list of boolean (flag) settings.
</fvec>
\end{lstlisting}

\begin{lstlisting}[language=XML]
<mvec name="PACKAGE:mvec" default="{0,1}" min="0" max="10">
This is an example variable length list of integer (mode) settings.
</mvec>
\end{lstlisting}

\begin{lstlisting}[language=XML]
<pvec name="PACKAGE:pvec" default="{0.0,1.0}" min="0.0" max="10.0">
This is an example variable length list of double-precision (parm) settings.
</pvec>
\end{lstlisting}

\begin{lstlisting}[language=XML]
<wvec name="PACKAGE:wvec" default="{value1,value2}">
This is an example variable length list of character-string (word) settings. Unlike for "word" these string values can contain spaces, but cannot contain commas, ",".
</wvec>
\end{lstlisting}

\hypertarget{examples}{%
\subsubsection{Examples}\label{examples}}

One of the best ways to demonstrate how to use code is through examples;
for context, \pythia has over 100 examples demonstrating various aspects
of the program. A very minimal example is provided via the template,
\vb{testPlugins.cc} located in
\vb{share/PACKAGE/examples}. The example creates a
\pythia object, reads any user-provided command files, and initializes
\pythia.

\begin{lstlisting}[language={C++}]
  Pythia pythia;
  pythia.readString("ProcessLevel:all = off");
  for (int iArg = 1; iArg < argc; ++iArg)
    pythia.readFile(argv[iArg]);
  pythia.init();
\end{lstlisting}

For each class with name \vb{NAME} created by the
template, a command file \vb{NAME.cmnd} is
provided.

\begin{lstlisting}
Init:plugins = {libPACKAGE.so::NAME}
\end{lstlisting}

The exceptions are the \vb{PhaseSpace},
\vb{SpaceShower}, and
\vb{TimeShower} classes, which cannot be passed to
\pythia through \vb{Init:plugins} on their own. For
these classes the test program \vb{testNAME.cc} is
provided instead, which loads the class with
\vb{make\_plugin}, and can be built with
\vb{make testNAME} in the examples directory.

In the \vb{MyPackage} example of the
\protect\hyperlink{test-the-package}{test the package} section, the test
can be run for the PDF class with the following.

\begin{lstlisting}[language=bash,style=shell]
./testPlugins MyPDF.cmnd
\end{lstlisting}

This will then load the \vb{MyPDF} class through
the plugin system.

A minimal build system is provided in the
\vb{examples} directory of the template. By default
it requires the plugin library to be built, and provides rules to build
any \vb{main*} programs as well as the
\vb{test*} programs. External packages can be
linked, for more details see \protect\hyperlink{build-system}{build
system}. The \vb{Makefile} first reads the
\vb{Makefile.inc} configuration in the same
directory, if it exists. It then tries to find the package's plugin
library \vb{libPACKAGE.so}, first looking in the
local path \vb{../../../lib} and then along the
\pythia library path, and stops with an error if it cannot be found.
Finally, it will build any examples with the following compiler flags.

\begin{itemize}
\tightlist
\item
  \vb{-Wl,-rpath,PACKAGE\_LIB\_PATH}: run-time path
  to the plugin library
\item
  \vb{-IPYTHIA8\_INCLUDE\_PATH}: path to the \pythia
  headers
\item
  \vb{-LPYTHIA8\_LIB\_PATH}: link-time path to the
  \pythia library
\item
  \vb{-Wl,-rpath,PYTHIA8\_LIB\_PATH}: run-time path
  to the \pythia library
\item
  \vb{-lpythia8}: the \pythia library
\end{itemize}

Note that the flags for the package headers path
(\vb{-IPACKAGE\_INCLUDE\_PATH}), the link-time path
to the plugin library (\vb{-LPACKAGE\_LIB\_PATH}),
and the plugin library (\vb{-lPACKAGE}) are not
passed to the compiler, as it is assumed that the examples load the
package's plugins through the \pythia plugin system, rather than using
the plugin classes directly. The run-time path to the plugin library
(\vb{-Wl,-rpath,PACKAGE\_LIB\_PATH}) must still be
specified to ensure that it can be found at run time.

\hypertarget{package-development}{%
\subsection{Package Development}\label{package-development}}

The \contrib system utilizes the \pythia plugin system. In short,
users request the use of a plugin by defining the
\vb{Init:plugins} setting.

\begin{lstlisting}
Init:plugins = {libPACKAGE1.so::NAME1,libPACKAGE2.so::NAME2,...}
\end{lstlisting}

Details on this setting syntax, as well as the \pythia plugin system, are
documented in the
\href{https://pythia.org/manuals/pythia8318/Plugins.html}{\pythia HTML
manual}. At its simplest, the plugin library of a \contrib package
consists of two parts: developer-defined plugin classes, derived from
the list of available base classes, and macros which allow these plugins
to be loaded by \pythia. These macros are summarized here, and are
explained in further detail below in
\protect\hyperlink{plugin-library-definition}{plugin library
definition}. These macros should not be used in header files.

\begin{itemize}
\tightlist
\item
  \vb{PYTHIA8\_PLUGIN\_CLASS}: registers a plugin
  class, called once per class.
\item
  \vb{PYTHIA8\_PLUGIN\_VERSIONS}: defines the
  compatible versions of \pythia, called once per plugin library.
\item
  \vb{PYTHIA8\_PLUGIN\_PARALLEL}: declares if the
  plugin library is compatible with
  \vb{PythiaParallel}, optionally called once per
  plugin library.
\item
  \vb{PYTHIA8\_PLUGIN\_SETTINGS}: registers
  settings for the plugin library, optionally called once per plugin
  library.
\item
  \vb{PYTHIA8\_PLUGIN\_XML}: defines XML settings
  for the plugin library, optionally called once per plugin library, see
  \protect\hyperlink{xml-settings}{XML settings} for further details.
\end{itemize}

If any of the four single-use macros are called more than once in a
package, the package will fail to compile.

\hypertarget{namespace}{%
\subsubsection{Namespace}\label{namespace}}

Because the \contrib system is intended for decentralized
development, it is very possible that there are conflicts in class names
between different packages. In general, this should not be an issue,
since the run-time loading mechanism of the \pythia plugin system
protects against this. However, if users wish to link against the plugin
library at build time, then this can become an issue. Consequently, it
is requested that package authors use the nested namespace
\vb{Pythia8::Pythia8Contrib::PACKAGE} for a package
with the name \vb{PACKAGE}. In the template header
and source, this convention is set with the following structure:

\begin{lstlisting}[language={C++}]
namespace Pythia8 {
namespace Pythia8Contrib {
namespace PACKAGE {
// PACKAGE code here.
} // end namespace PACKAGE
} // end namespace Pythia8Contrib
} // end namespace Pythia8
\end{lstlisting}

Without \vb{using} or
\vb{namespace} statements, this means that
accessing objects can be somewhat cumbersome,

\begin{lstlisting}[language={C++}]
Pythia8::Pythia8Contrib::PACKAGE::NAME
\end{lstlisting}

where \vb{NAME} is the object being accessed.
Consequently, it is suggested that the above
\vb{namespace} code is always used to enclose the
body of header and source files. Alternatively,
\vb{using <NAMESPACE>} statements can be used
instead, although this is not recommended; these statements should only
be used in source files (\vb{*.cc}) and never
included in header files (\vb{*.h}).

\hypertarget{headers}{%
\subsubsection{Headers}\label{headers}}

The headers produced by the template generator are all located in
\vb{include/PACKAGE}. One header,
\vb{include/PACKAGE/NAME.h}, is generated for each
\vb{BASE/NAME} pair passed as an argument to
\vb{generate}. This header is intended to define
the structure of the class, without including the implementation, which
is intended for the corresponding source file,
\vb{src/NAME.cc}. However, in some cases it may be
necessary to provide implementations in the header, \emph{e.g.}, for the
case of templated classes or methods. In general, the macros summarized
at the start of this section should not be used in the headers.

When a package is installed with the \contrib system, the headers
will be placed in the location
\vb{include/Pythia8Contrib/PACKAGE}, \emph{i.e.},
with an additional level of nesting. This should not cause issues, since
an additional include path can always be passed to the compiler,
\vb{-Iinclude/Pythia8Contrib}. In this way the
package headers should always be included with statements such as,

\begin{lstlisting}[language={C++}]
#include "PACKAGE/NAME.h"
\end{lstlisting}

without the need for prefixing the include statement with
\vb{Pythia8Contrib/}. This is what is done in the
\vb{Makefile} for the examples located in
\vb{share/PACKAGE/examples}, where the location of
the package is automatically determined to either be local or part of a
\pythia installation.

The constructor for the class \vb{NAME} defined by
the header \vb{NAME.h} must take the following
form:

\begin{lstlisting}[language={C++}]
NAME(Pythia* pythiaPtr, Settings* settingsPtr, Logger* loggerPtr) 
\end{lstlisting}

There are three pointers which may be passed to the constructor:

\begin{itemize}
\tightlist
\item
  \vb{pythiaPtr}: of type
  \vb{Pythia*}. This provides access to the full
  \pythia generator object which is loading the plugin class.
\item
  \vb{settingsPtr}: of type
  \vb{Settings*}. If the entire \pythia generator
  object is not passed, this provides access to a settings object.
\item
  \vb{loggerPtr}: of type
  \vb{Logger*}. Again, if the entire \pythia
  generator is not passed, this provides access to a logging service.
\end{itemize}

When a class is constructed from a \pythia object as a plugin,
\emph{i.e.}, from the \vb{Init:plugins} setting,
all three pointers are valid. However, it is also possible for plugins
to be created directly in the C++ code, where none of these pointers is
required to be valid; further details are given in the following
section.

When using the full template produced with
\vb{generate}, all required methods for the class
will be included in the corresponding header. For example, the
\vb{UserHooks} base class does not have any pure
virtual methods, and so other than the constructor and destructor, no
additional methods need be implemented. However, there are a large
number of methods that can be overridden from the base class. In the
template, each optionally implemented method will be included as a
commented-out block of code.

\begin{lstlisting}[language={C++}]
  // // Initialisation after beams have been set by Pythia::init().
  // bool initAfterBeams() override
  // {return true;}
\end{lstlisting}

Here, code is given which will override the
\vb{initAfterBeams} method of the base
\vb{UserHooks} class and return the value of
\vb{true}. The general form of the comment will
contain the following three sections: a commented description of the
method, the call signature of the method, and optionally the
implementation of the method if sufficiently succinct for a header. Note
that these methods will always end with the
\vb{override} keyword to indicate that this method
should be overriding the base-class implementation.

An example of a required method can be found in a
\vb{PDF} derived class where the protected method
\vb{xfUpdate} must be implemented.

\begin{lstlisting}[language={C++}]
  // Update parton densities (defined in src/MyPDF.cc). Must be defined.
  void xfUpdate(int id, double x, double Q2) override;
\end{lstlisting}

Note that here no implementation is given, which is instead relegated to
the source file.

\hypertarget{source}{%
\subsubsection{Source}\label{source}}

For each developer-defined class with name
\vb{NAME}, there is a corresponding
\vb{src/NAME.cc} source file. This file is intended
to include any detailed implementations that are not suitable for the
corresponding \vb{include/PACKAGE/NAME.h} header.
In the generated template, an implementation for the constructor is
always provided, which demonstrates how to use methods of the passed
\vb{Pythia} pointer, access settings from the
passed \vb{Settings} pointer, and write messages
via the passed \vb{Logger} pointer. The following
provides an abbreviated template for derived class
\vb{NAME} of base class
\vb{BASE}.

\begin{lstlisting}[language={C++}]
//--------------------------------------------------------------------------

// Constructor.

NAME::NAME(
  Pythia* pythiaPtrIn, Settings* settingsPtrIn,
  Logger* loggerPtrIn) : BASE() {

  // If a pointer is not needed its name (but not type) can be
  // commented out above, e.g. change pythiaPtrIn to /*pythiaPtrIn*/,
  // to prevent compiler warnings. If a pointer is not needed, ensure
  // it is specified as not required in the corresponding call to the
  // macro PYTHIA8_PLUGIN_CLASS.
  
  // Include whatever code is needed here to construct the plugin
  // class. Below, we give some examples of how the pointer arguments
  // can be used.

  // Access and print the settings as defined by the XML files.
  loggerPtrIn->INFO_MSG("XML setting PACKAGE:flag is " + to_string(
      settingsPtrIn->flag("PACKAGE:flag")));
  // ...
  
}
\end{lstlisting}

As discussed in the \protect\hyperlink{headers}{headers} section, the
passed pointers may not be required to be valid. This is specified by
the macro \vb{PYTHIA8\_PLUGIN\_CLASS}, which
registers the class so that it can be accessed from the plugin library.
The usage of \vb{PYTHIA8\_PLUGIN\_CLASS} is as
follows.

\begin{lstlisting}[language={C++}]
PYTHIA8_PLUGIN_CLASS(BASE, NAME,
  true, // Require the Pythia pointer to be valid.
  true, // Require the Settings pointer to be valid.
  true) // Require the Logger pointer to be valid.
\end{lstlisting}

Here, \vb{BASE} is the base class, \emph{e.g.},
\vb{UserHooks}, while
\vb{NAME} is the derived class to be registered,
\emph{e.g.}, \vb{MyUserHooks}. If an argument is
\vb{true} then that type of pointer must be a valid
pointer when using the \pythia plugin system to load the plugin. If
\vb{false}, then that pointer may be passed as a
null pointer. If a pointer is specified by the package author as not
being required, then it is important that the validity of that pointer
is checked in the implemented constructor for the derived class.

In the \pythia plugin system, plugins may be loaded either through
settings or through using \vb{make\_plugin} in
code, with four different possible use cases:

\begin{enumerate}
\def\labelenumi{\arabic{enumi}.}
\item
  Via user-defined settings, using the format
  \vb{Init:plugins = \{libPACKAGE.so::NAME\}}.
  Here, all three pointers passed will always be valid, and correspond
  to the \vb{Pythia} object loading the plugin.
\item
  Via the following \vb{make\_plugin} method called
  in C++, where no pointers are passed:

\begin{lstlisting}
 shared_ptr<BASE> make_plugin<BASE>(string PACKAGE, string NAME)
\end{lstlisting}

  Here, the \vb{PACKAGE} string would be given as
  \vb{libPACKAGE.so} and
  \vb{NAME} is the name of the plugin class to be
  loaded. All three pointers loaded with this method are null. If any of
  the \vb{PYTHIA8\_PLUGIN\_CLASS} pointer arguments
  are set as \vb{true}, then this method will fail
  and return a null pointer.
\item
  Via the following \vb{make\_plugin} method called
  in C++, where now a \vb{Pythia} pointer is
  passed:

\begin{lstlisting}
 shared_ptr<BASE> make_plugin<BASE>(string PACKAGE, string NAME, Pythia* pythiaPtr)
\end{lstlisting}

  If the \vb{Pythia} pointer is valid, then all
  three pointers to the plugin constructor will be valid and taken from
  the \vb{Pythia} pointer. If the
  \vb{Pythia} pointer is null and any of the
  pointers are required, then this method will fail and return a null
  pointer.
\item
  Via the following \vb{make\_plugin} method called
  in C++, where all three pointers are passed individually:

\begin{lstlisting}
 shared_ptr<BASE> make_plugin<BASE>(string PACKAGE, string NAME, Pythia* pythiaPtr, Settings* settingsPtr, Logger* loggerPtr)
\end{lstlisting}

  If \vb{pythiaPtr} is valid, and either
  \vb{settingsPtr} or
  \vb{loggerPtr} is null, then the corresponding
  pointer from the \vb{pythiaPtr} will be passed to
  the plugin class constructor instead. If any of the pointers are
  required by the \vb{PYTHIA8\_PLUGIN\_CLASS} call,
  and that corresponding pointer is null, then this method will fail and
  return a null pointer.
\end{enumerate}

In addition to the required implementation of the constructor, a number
of methods may be required, dependent upon the base class. The generated
template source file for each developer-defined class will contain an
outline of how to implement the required methods. In the example above
for the \vb{PDF} derived class, an implementation
of the required \vb{xfUpdate} method is given.

\begin{lstlisting}[language={C++}]
//--------------------------------------------------------------------------

// Calculates the PDF distributions for the specified parton id at (x,
// Q^2) and stores the resulting values in the corresponding
// fields. If id == 9, the values should be calculated for all
// relevant partons. The user may choose to always calculate the value
// for all partons, in which case they should set idSav = 9. The PDF
// class flexibly handles antiparticle and isospin symmetries, so
// xfUpdate should always give the values that correspond to the
// particle with positive id and isospin.

void NAME::xfUpdate(int /*id*/, double /*x*/, double /*Q2*/) {

  // // Update the values (currently 0 is used as a dummy value).
  // xg     = 0;
  // ...

  // // idSav = 9 to indicate that all flavours reset, otherwise use id.
  // idSav = 9;
  
}
\end{lstlisting}

Optional methods with more lengthy implementations are also provided in
the template source file. For a \vb{UserHooks}
derived class, the following commented-out code implements the optional
\vb{multiplySigmaBy} method:

\begin{lstlisting}[language={C++}]
//--------------------------------------------------------------------------

// multiplySigmaBy allows the user to introduce a multiplicative factor
// that modifies the cross section of a hard process. Since it is called
// from before the event record is generated in full, the normal analysis
// does not work. The code here provides a rather extensive summary of
// which methods actually do work. It is a convenient starting point for
// writing your own derived routine.

// double NAME::multiplySigmaBy(const SigmaProcess* sigmaProcessPtr,
//   const PhaseSpace* phaseSpacePtr, bool inEvent) {
// 
//   // Process code, necessary when some to be treated differently.
//   int code       = sigmaProcessPtr->code();
// 
//   // Final multiplicity, i.e. whether 2 -> 1 or 2 -> 2.
//   int nFinal     = sigmaProcessPtr->nFinal();
//
//   // Incoming x1 and x2 to the hard collision, and factorization scale.
//   double x1      = phaseSpacePtr->x1();
//   double x2      = phaseSpacePtr->x2();
//   double Q2Fac   = sigmaProcessPtr->Q2Fac();
//   // ...
//
//   // Dummy statement.
//   return 1.;
//
// }
\end{lstlisting}

This commented code provides a brief outline of how the method could be
implemented, including access of relevant data members from the class.

\hypertarget{thread-safety}{%
\subsubsection{Thread Safety}\label{thread-safety}}

By default, plugin classes are assumed to be thread-unsafe to prevent
difficult-to-debug errors when running in a threaded environment.
Consequently, plugin classes cannot be loaded by
\vb{PythiaParallel} unless the
\vb{PYTHIA8\_PLUGIN\_PARALLEL} macro is called with
\vb{true} for a plugin library, see the
\protect\hyperlink{plugin-library-definition}{plugin library definition}
section for details. In some cases, plugin classes may need to lock a
thread to prevent data races on shared resources, \emph{e.g.},
histograms. To enable this, every class derived from
\vb{PhysicsBase} has access to
\vb{mutexPtr}, a mutex pointer shared across all
threads of the top-level \vb{PythiaParallel}
instance. Whenever this mutex is locked, the following code pattern will
prevent threading:

\begin{lstlisting}[language={C++}]
mutexPtr->lock();
// Thread unsafe code can be written here.
mutexPtr->unlock();
// Thread safe code can be written here.
\end{lstlisting}

All classes that can be passed to a \vb{Pythia}
instance by an external pointer are derived from
\vb{PhysicsBase} and have access to this
\vb{mutexPtr} except \vb{PDF},
\vb{BeamShape}, and
\vb{RndmEngine}.

For many plugin classes that are threaded, it is necessary to combine
the output of threads of \vb{PythiaParallel}
together again. This can be accomplished for classes inheriting from
\vb{PhysicsBase} by defining
\vb{onStat}.

\begin{lstlisting}[language={C++}]
void NAME::onStat(vector<PhysicsBase*> pluginPtrs, Pythia* pythiaPtrIn) { 
  // Loop over the plugin from each thread. 
  for (PhysicsBase* pluginPtr : pluginPtrs) { 
    // Skip the plugin that this method was called for. 
    if (pluginPtr == this) continue; 
    // Access all other plugins. 
    NAME* pluginNow = dynamic_cast<NAME*>(pluginPtr); 
    // Perform whatever merging is necessary here. 
  } 
  // Call onStat for this plugin-class instance. 
  onStat(); 
}
\end{lstlisting}

Here, \vb{pluginPtrs} is a vector of the
plugin-class pointer from each thread. Each of these plugin-class
pointers then needs to be recast back to the plugin class. Finally, the
\vb{onStat()} for this instance of the plugin
should be called, with no arguments. This ensures that the
\vb{onStat} behavior is the same between a single
\vb{Pythia} instance, and
\vb{PythiaParallel}. The
\vb{onStat} method defined above is called when
\vb{PythiaParallel::stat()} is called.

\hypertarget{plugin-library-definition}{%
\subsubsection{Plugin Library
Definition}\label{plugin-library-definition}}

The remaining four macros given at the introduction of this
\protect\hyperlink{package-development}{package development} section can
only be called once per plugin library, unlike
\vb{PYTHIA8\_PLUGIN\_CLASS} which must be called
per class definition. If called more than once in the package code, the
package will fail to compile. These macros define properties of the
plugin library and can be called anywhere in the source files. However,
it is preferred that they are all called in a single source file, which
in the generated template is done in
\vb{src/Package.cc}. The first macro,

\begin{lstlisting}[language={C++}]
PYTHIA8_PLUGIN_VERSIONS(8310, 8311, ...)
\end{lstlisting}

defines the versions of \pythia that the plugin library is compatible
with. Any number of versions can be passed. When a plugin library is
loaded by \pythia, a check is made to see if the version of \pythia
loading the plugin library is compatible. If it is not, a warning is
issued, and a null pointer is returned for any class being constructed
from that plugin library.

The second macro,

\begin{lstlisting}[language={C++}]
PYTHIA8_PLUGIN_PARALLEL(true)
\end{lstlisting}

is optional, and declares if the plugin library is compatible with
\vb{PythiaParallel}. If not called, the plugin
library is assumed to be incompatible with
\vb{PythiaParallel} which is the same as calling
\vb{PYTHIA8\_PLUGIN\_PARALLEL} with an argument of
\vb{false}. See
\protect\hyperlink{thread-safety}{thread safety} for what must be true
of the plugin classes before this can be set to
\vb{true}.

The third macro,

\begin{lstlisting}[language={C++}]
PYTHIA8_PLUGIN_XML("PACKAGE/xmldoc/Index.xml") 
\end{lstlisting}

is optional, but if called, requires XML files for the plugin library to
be present. Any starting index file can be passed, but in the template
the file \vb{PACKAGE/xmldoc/Index.xml} is used as
the convention. More details on how settings can be defined via XML, as
well as the search path for the starting XML file are given in the
\protect\hyperlink{xml-settings}{XML settings} section of this
documentation.

Finally, settings can be registered directly in the C++ code, rather
than via XML files, through interactions with a
\vb{Settings} object with the following macro:

\begin{lstlisting}[language={C++}]
PYTHIA8_PLUGIN_SETTINGS(PACKAGESettings)
\end{lstlisting}

The method \vb{PACKAGESettings} can take any name,
but must be defined with the following call signature:

\begin{lstlisting}[language={C++}]
void PACKAGESettings(Settings* settingsPtr)
\end{lstlisting}

When this method is called, \vb{settingsPtr} is
guaranteed to be valid and does not need to be checked. The
\vb{PACKAGESettings} method is always called after
XML settings are loaded.

The following \vb{add} methods for the
\vb{Settings} class allow settings to be added to
the settings database. Possible setting types are boolean
(\vb{flag}), integer
(\vb{mode}), double
(\vb{parm}), string
(\vb{word}), vector of booleans
(\vb{fvec}), vector of integers
(\vb{mvec}), vector of doubles
(\vb{pvec}), and vector of strings
(\vb{wvec}). It is recommended that all settings
begin with the \vb{PACKAGE} prefix, to prevent
collisions with settings from other \contrib packages, as well as
\pythia itself. The arguments to these methods are as follows:

\begin{itemize}
\tightlist
\item
  \vb{keyIn}: a string that is the setting name,
  \emph{i.e.}, when reading a command string such as
  \vb{MySetting = Value},
  \vb{keyIn} is
  \vb{MySetting}.
\item
  \vb{defaultIn}: default value for the setting,
  must be of the same type as the \vb{add} method
  called, \emph{e.g.}, the \vb{defaultIn} for
  \vb{addMode} is an integer.
\item
  \vb{hasMinIn}: boolean flag, if there is a
  minimum allowed value for the setting.
\item
  \vb{hasMaxIn}: same as
  \vb{hasMinIn}, but for maximum.
\item
  \vb{minIn}: specifies the optional minimum, of
  the same type as the method, \emph{e.g.}, a double for
  \vb{addParm}.
\item
  \vb{maxIn}: same as
  \vb{minIn}, but for maximum.
\item
  \vb{optOnlyIn}: a boolean which specifies only
  allowed options are considered valid, for the
  \vb{addMode} method only.
\end{itemize}

The methods themselves are:

\begin{lstlisting}[language={C++}]
// Add a boolean setting (flag).
settingsPtr->addFlag(string keyIn, bool defaultIn);
\end{lstlisting}

\begin{lstlisting}[language={C++}]
// Add an integer setting (mode).
settingsPtr->addMode(string keyIn, int defaultIn, bool hasMinIn,
  bool hasMaxIn, int minIn, int maxIn, bool optOnlyIn = false);
\end{lstlisting}

\begin{lstlisting}[language={C++}]
// Add a double setting (parm).
settingsPtr->addParm(string keyIn, double defaultIn, bool hasMinIn,
  bool hasMaxIn, double minIn, double maxIn);
\end{lstlisting}

\begin{lstlisting}[language={C++}]
// Add a string setting (word).
settingsPtr->addWord(string keyIn, string defaultIn);
\end{lstlisting}

\begin{lstlisting}[language={C++}]
// Add a vector of booleans setting (fvec).
settingsPtr->addFVec(string keyIn, vector<bool> defaultIn);
\end{lstlisting}

\begin{lstlisting}[language={C++}]
// Add a vector of integers setting (mvec).
settingsPtr->addMVec(string keyIn, vector<int> defaultIn, bool hasMinIn,
  bool hasMaxIn, int minIn, int maxIn);
\end{lstlisting}

\begin{lstlisting}[language={C++}]
// Add a vector of doubles setting (pvec).
settingsPtr->addPVec(string keyIn, vector<double> defaultIn, bool hasMinIn,
  bool hasMaxIn, double minIn, double maxIn);
\end{lstlisting}

\begin{lstlisting}[language={C++}]
// Add a vector of strings setting (wvec).
settingsPtr->addWVec(string keyIn, vector<string> defaultIn);
\end{lstlisting}

\hypertarget{formatting}{%
\subsubsection{Formatting}\label{formatting}}

No formatting requirements are defined for \contrib packages,
although it is recommended to follow the \pythia formatting style. The
generated templates follow the \pythia coding style, which is provided in
the somewhat lengthy
\href{https://gitlab.com/Pythia8/releases/-/blob/pythia8318/CODINGSTYLE}{\vb{CODINGSTYLE}}
file distributed with the \pythia source. For convenience a few points
are summarized here.

\begin{itemize}
\item
  Use existing files to get a feel for the general style.
\item
  English is the only allowed language (for comments, variable names,
  \emph{etc.}).
\item
  Lines should be at most 79 characters long, so that they do not
  overflow when opened in an 80-character-wide text editor window.
\item
  The underscore character, \vb{\_}, should be
  avoided as far as possible.
\item
  Use lines

\begin{lstlisting}
  //==========================================================================
\end{lstlisting}

  to separate classes from each other, and from top and bottom material
  of a file, that does not belong to a class.
\item
  Use lines

\begin{lstlisting}
  //--------------------------------------------------------------------------
\end{lstlisting}

  for smaller subdivisions than above. Specifically, in
  \vb{.cc} files, insert it between the different
  methods that belong to the same class.
\item
  Do not use tabs.
\item
  Do not use comments of the style \vb{/* .... */}
  except when needed to clarify an argument name that is not used in a
  call signature.
\item
  Use the \pythia types when available.
\end{itemize}

\hypertarget{build-system}{%
\subsection{Build System}\label{build-system}}

There are any number of build system tools available, and in many cases
package authors have strong preferences for certain systems over others.
Consequently, there are no restrictions on what build systems package
authors use for their package. However, in many cases, package authors
may want a build system out-of-the-box, and so we provide such a build
system based on the \pythia build system. This build system has
relatively few dependencies, other than standard POSIX tools. First we
describe the interaction that the build system (whatever that system is)
must have with \contrib, then we describe how to customize the
provided build system in more detail.

\hypertarget{pythia-contrib-interaction}{%
\subsubsection{\contrib
Interaction}\label{pythia-contrib-interaction}}

The top-level build system for \contrib consists of a
\vb{configure} script and a
\vb{Makefile}. Each \contrib package is then
expected to consist of a folder in the top level of \contrib. The
general build procedure for \contrib is as follows:

\begin{lstlisting}[language=bash,style=shell]
./configure [OPTIONS]
make
make install
\end{lstlisting}

This will then configure and build all packages in the top level of
\contrib. It is expected that the package-provided build system
optionally has a \vb{configure} script, and must
have a \vb{Makefile}. Further specifications for
each of these components are given below.

\hypertarget{configure-requirements}{%
\paragraph{Configure Requirements}\label{configure-requirements}}

The top-level \contrib \vb{configure} script
is designed so that users can configure all the \contrib packages
at the top level without needing to configure each package individually.
To this end, the top-level \vb{configure} script
takes options of the form:

\begin{lstlisting}
--with-DEP[=DIR]       : Use the external dependency DEP  with
                         its top directory optionally set by DIR. If the
                         enabled DEP requires other external packages,
                         these packages will also be enabled.
--with-DEP-bin=DIR     : Set the path to the DEP binaries.
--with-DEP-lib=DIR     : Set the path to the DEP libraries.
--with-DEP-include=DIR : Set the path to the DEP headers.
--with-DEP-config=BIN  : Set the package configuration script BIN.
\end{lstlisting}

and passes these on to the \vb{configure} script
for each package. Note that here, \vb{DEP} should
be passed in all lower case.

Additionally, the option \vb{--nowarn} is passed,
which tells the package-level \vb{configure} script
to not issue warnings for \vb{--with} options that
are not available. In this way, if package \emph{A} requires the
dependency \vb{fastjet3}, while package \emph{B}
does not, package \emph{B} will not warn that it is being passed an
invalid option \vb{--with-fastjet3}. The top-level
\vb{configure} will skip calling the package-level
\vb{configure} script if it is missing, and will
issue a warning instead that the package must be configured manually,
following documentation that should be provided in the package
\vb{README.md}.

It is expected that if a package is configured, then a package-level
\vb{Makefile.inc} will be present. This file may be
empty if a custom build system is being used, but is intended to be
included by the package-level \vb{Makefile}. The
\contrib \vb{configure} script will skip
running the package-level \vb{configure} if the
package-level \vb{Makefile.inc} is present, unless
the \vb{--force} option is passed.

To summarize, the package-level \vb{configure}
script is optional. If it does exist, it should adhere to the following
specifications:

\begin{itemize}
\tightlist
\item
  Accept \vb{--with-DEP} options with variations
  \vb{-bin}, \vb{-lib},
  \vb{-include}, and
  \vb{-config}, as described above, for dependency
  \vb{DEP}.
\item
  Accept the \vb{--nowarn} option, which turns off
  any warnings that might be generated when passing
  \vb{--with-DEP} options.
\item
  Generate a package-level \vb{Makefile.inc} file
  after being run. This file may be empty, but is intended to be
  included by the package-level \vb{Makefile}.
\end{itemize}

\hypertarget{make-requirements}{%
\paragraph{Make Requirements}\label{make-requirements}}

The top-level \contrib build system uses a
\vb{Makefile} to build all the packages. It does
this by calling

\begin{lstlisting}
make -C PACKAGE contrib
\end{lstlisting}

for each package with name \vb{PACKAGE}. The
package-level \vb{Makefile} should then perform the
following actions when called with target
\vb{contrib}:

\begin{itemize}
\tightlist
\item
  Produce the plugin library \vb{lib/libPACKAGE.so}
  at the top level of \contrib.
\item
  Optionally, copy relevant data and documentation to \contrib
  \vb{share/Pythia8Contrib/PACKAGE}.
\item
  Optionally, copy relevant header files to \contrib
  \vb{include/Pythia8Contrib/PACKAGE}.
\item
  Optionally, copy relevant executables to \contrib
  \vb{bin}.
\end{itemize}

If the library \vb{lib/libPACKAGE.so} is not
produced, then it is assumed that the build has failed. Note that the
\vb{.so} suffix does not change between build
platforms, \emph{e.g.}, the \vb{.dylib} suffix
common to macOS is not used. This is because plugin libraries are
intended to be loaded dynamically at run time as set by configuration
files or strings, and so to ensure that the configuration files are not
platform-dependent, the same library suffix is used across all
platforms.

\hypertarget{configuration-customization}{%
\subsubsection{Configuration
Customization}\label{configuration-customization}}

When using the provided build system, the following settings with the
syntax \vb{KEY=VALUE} can be used to configure
external packages in the optional
\vb{.configure.inc} file provided at the top level
of the package. It is possible to define multiple external packages in
\vb{.configure.inc}, and by default a number of
common packages are defined in the template
\vb{.configure.inc}. Each external package
configuration starts with a line beginning with
\vb{PKG=}. If options for a package are missing,
then they are taken as the default value. Note, each configuration key
must begin the line and follow the syntax
\vb{KEY=VALUE}. The following keys are allowed
(defaults, if defined, are given in brackets after the key):

\begin{itemize}
\tightlist
\item
  \vb{PKG}: name of the external package to
  configure (case does not matter). The configuration script will then
  use this name for configuring the package. For example,
  \vb{PKG=LHAPDF6} will then include the options
  \vb{--with-lhapdf6}.
\item
  \vb{CFG}: optional configure script for the
  package, for example \vb{fastjet-config}.
\item
  \vb{BIN\_PATH}
  {[}\vb{bin}{]}: given the root package directory,
  relative path to the binary directory, or absolute path. By default,
  this is the relative path \vb{bin}.
  Alternatively, this can be an argument given to the configure script
  specified by \vb{CFG}. For example, this could be
  set to \vb{--bindir}, if the configure script
  accepts this as a valid argument. If the configure script is not
  available, or if this argument is invalid, the default value of
  \vb{bin} will be used.
\item
  \vb{INC\_PATH}
  {[}\vb{include}{]}: same as for
  \vb{BIN\_PATH}, but for the header directory.
\item
  \vb{LIB\_PATH}
  {[}\vb{lib}{]}: same as above but for the library
  directory.
\item
  \vb{BIN\_DEPS}: required executables along
  \vb{BIN\_PATH}, given as a semi-colon separated
  list. For example, \vb{rootcint;root-config}
  requires the executables \vb{rootcint} and
  \vb{root-config} exist along
  \vb{BIN\_PATH}.
\item
  \vb{INC\_DEPS}: same as above for
  \vb{BIN\_DEPS}, except now required headers along
  \vb{INC\_PATH}.
\item
  \vb{LIB\_DEPS}: same as above for
  \vb{BIN\_DEPS}, except now required libraries
  along \vb{LIB\_PATH}.
\item
  \vb{PKG\_DEPS}: it is possible to require
  additional external packages as dependencies for this external
  package. This is typically needed when there is no configure script
  for the external package. The list should be provided as a semi-colon
  separated list, for example
  \vb{LHAPDF6;FASTJET3}. These packages then should
  also be defined in this configuration.
\item
  \vb{FLAGS}: optionally specify any additional
  compiler flags that should be used when this package is enabled. For
  example, preprocessor directives can be passed with the option
  \vb{-D<argument>}.
\item
  \vb{REQUIRE}
  {[}\vb{false}{]}: if set to
  \vb{true}, \emph{e.g.},
  \vb{REQUIRE=true}, this dependency is required,
  and the configuration will fail unless the package is configured.
\end{itemize}

By default, a number of common external dependencies are fully defined
in \vb{.configure.inc}. It is suggested that the
following names are used for these common dependencies (links to the
corresponding website for each dependency are also given):

\begin{itemize}
\tightlist
\item
  \href{https://pythia.org/}{\vb{PYTHIA8}}\cite{Bierlich:2022pfr}: a
  program for generating high-energy physics collision events,
  \emph{i.e.}, for the description of collisions at high energies
  between electrons, protons, photons and heavy nuclei.
\item
  \href{https://evtgen.hepforge.org/}{\vb{EVTGEN}}\cite{Lange:2001uf}:
  a Monte Carlo event generator that simulates the decays of
  heavy-flavor particles, primarily \(B\) and \(D\) mesons.
\item
  \href{https://fastjet.fr/}{\vb{FASTJET3}}\cite{Cacciari:2011ma}: a
  software package for jet finding in \(pp\) and \(e^+e^-\) collisions.
\item
  \href{https://gitlab.cern.ch/hepmc/HepMC}{\vb{HEPMC2}}\cite{Dobbs:2001ck}:
  a framework for Monte Carlo event record encoding and manipulation.
\item
  \href{https://gitlab.cern.ch/hepmc/HepMC3}{\vb{HEPMC3}}\cite{Buckley:2019xhk}:
  the next-generation framework of \vb{HEPMC},
  recommended over \vb{HEPMC2}.
\item
  \href{https://lhapdf.hepforge.org/lhapdf5/}{\vb{LHAPDF5}}\cite{Whalley:2005nh}:
  a unified and easy-to-use interface to modern PDF sets.
\item
  \href{https://lhapdf.hepforge.org/}{\vb{LHAPDF6}}\cite{Buckley:2014ana}:
  the next-generation version of \vb{LHAPDF},
  recommended over \vb{LHAPDF5}.
\item
  \href{https://rivet.hepforge.org/}{\vb{RIVET}}\cite{Bierlich:2024vqo}: a
  system for validation of Monte Carlo event generators (Robust
  Independent Validation of Experiment and Theory).
\item
  \href{https://root.cern.ch/}{\vb{ROOT}}\cite{Brun:1997pa}: an
  open-source data analysis framework used by high-energy physics and
  others.
\item
  \href{https://www.gnu.org/software/gzip/}{\vb{GZIP}}:
  a popular data compression program.
\item
  \href{https://www.python.org/}{\vb{PYTHON}}: a
  general-purpose programming language.
\item
  \href{https://www.openmp.org/}{\vb{OPENMP}}: a
  portable, scalable model that gives shared-memory parallel programmers
  a simple and flexible interface for developing parallel applications.
\item
  \href{https://www.mpich.org/}{\vb{MPICH}}: a
  high-performance and widely portable implementation of the message
  passing interface.
\item
  \href{https://www.hdfgroup.org/solutions/hdf5/}{\vb{HDF5}}:
  a high-performance data management and storage suite.
\item
  \href{https://github.com/BlueBrain/HighFive}{\vb{HIGHFIVE}}:
  a modern header-only \vb{C++14}-friendly
  interface for \vb{HDF5}.
\end{itemize}

\hypertarget{make-customization}{%
\subsubsection{Make Customization}\label{make-customization}}

For each external dependency specified with the
\vb{configure} script, a number of variables are
available within the \vb{Makefile} both at the top
level of the package, as well as in
\vb{share/PACKAGE/examples}. The defined variables
are as follows. Note that, other than
\vb{DEP\_USE}, these variables can be empty -- when
either the dependency has not been configured, or if the dependency does
not require that variable.

\begin{itemize}
\tightlist
\item
  \vb{DEP\_USE}: \vb{true}, if
  the package is successfully enabled, \vb{false}
  otherwise.
\item
  \vb{DEP\_CONFIG}: the configure script.
\item
  \vb{DEP\_BIN}: the path to the binary directory
  for the package.
\item
  \vb{DEP\_INCLUDE}: the header include flags.
\item
  \vb{DEP\_LIB}: the library flags used for
  linking, both paths and libraries.
\end{itemize}

Here, replace \vb{DEP} with (in all caps) the name
of the relevant dependency, \emph{e.g.},
\vb{PYTHIA8}.

The following lines in \vb{.configure.inc} define
and require \pythia as a dependency:

\begin{lstlisting}
PKG=PYTHIA8
CFG=pythia8-config
INC_PATH=--cxxflags
LIB_PATH=--libs
INC_DEPS=Pythia8/Plugins.h
REQUIRE=true
\end{lstlisting}

As an example, consider a \pythia installation at
\vb{/hep/pythia/8.3XX}. By running
\vb{./configure}, and assuming
\vb{/hep/pythia/8.3XX/bin/pythia8-config} is along
\vb{PATH}, the following lines will be generated in
the \vb{Makefile.inc}:

\begin{lstlisting}
PYTHIA8_USE=true
PYTHIA8_CONFIG=pythia8-config
PYTHIA8_BIN=/hep/pythia/8.3XX/bin/
PYTHIA8_INCLUDE=-I/hep/pythia/8.3XX/include
PYTHIA8_LIB=-L/hep/pythia/8.3XX/lib -Wl,-rpath,/hep/pythia/8.3XX/lib
\end{lstlisting}

Note that here the \pythia library is not included as part of the
dependency. However, if the line,

\begin{lstlisting}
LIB_DEPS=pythia8
\end{lstlisting}

is added to \vb{.configure.inc} then the
\vb{PYTHIA8\_LIB} variable will now include
\vb{-lpythia8}.

\begin{lstlisting}
PYTHIA8_LIB=-L/hep/pythia/8.3XX/lib -Wl,-rpath,/hep/pythia/8.3XX/lib -lpythia8
\end{lstlisting}

Assuming a dependency has been defined, it is then possible to modify
the \vb{Makefile} to link appropriately with the
dependency.

\begin{lstlisting}[language=make]
<target>: <prerequisites> 
ifeq ($(DEP_USE),true)
    <recipe, given the package exists>
else
    <recipe, given the package does not exist, or, error message>
    $(error Error: $@ requires DEP)
endif
\end{lstlisting}

Here, \vb{DEP} should be changed to whatever
dependency is being required. For the specific example of linking the
plugin library against a dependency, the plugin library rule would be
changed as follows:

\begin{lstlisting}[language=make]
ifeq ($(DEP_USE),true)
$(LOCAL_LIB)/libPACKAGE.so: $(OBJECTS)
    $(CXX) $^ -o $@ $(CXX_COMMON) $(CXX_SHARED) $(CXX_SONAME)$(notdir $@)\
     $(DEP_INCLUDE) $(DEP_LIB)
else
     $(error Error: $@ requires DEP)
endif
\end{lstlisting}

\hypertarget{package-submission}{%
\subsection{Package Submission}\label{package-submission}}

Packages for \contrib are recommended to be installed by users
through the
\href{https://gitlab.com/pythia8-contrib/use}{\vb{use}}
repository. The abbreviated package installation procedure is as
follows:

\begin{lstlisting}[language=bash,style=shell]
./enable PACKAGE[/VERSION] ... [OPTIONS]
./configure [OPTIONS]
make
make install
\end{lstlisting}

Here, \vb{PACKAGE} is the package name, followed by
an optional package version, and additional options which can be passed.
If no version is given, then the latest version is taken. For a package
to be available via the \vb{enable} command from
\contrib, it must have a repository as part of the
\href{https://gitlab.com/pythia8-contrib/packages}{\vb{packages}}
subgroup within \contrib. To request a repository for a new
package, please open an issue on the
\href{https://gitlab.com/pythia8-contrib/develop/-/issues}{\vb{develop}}
repository. The creation of this issue requires a freely available
\vb{gitlab.com} account. The issue should contain
the following information:

\begin{itemize}
\tightlist
\item
  requested name of the package,
\item
  brief description of the package.
\end{itemize}

\hypertarget{package-naming}{%
\subsubsection{Package Naming}\label{package-naming}}

Before creating a package, it is recommended to first check which names
are already used in the
\href{https://gitlab.com/pythia8-contrib/packages}{\vb{packages}}
subgroup of \contrib. While no strict naming restrictions are
enforced, the following conventions are encouraged:

\begin{itemize}
\tightlist
\item
  Only use letters (\emph{i.e.}, no numbers or special characters such
  as \vb{+}, spaces, \vb{-},
  \vb{\_}, \emph{etc.}). Special characters such as
  spaces or \vb{+} can easily break build systems
  while numbers in general should be reserved for versioning tags.
\item
  Begin the package name with a capital letter, and capitalize the first
  letter of each subsequent word, \emph{e.g.},
  \vb{MyPackageName}.
\end{itemize}

\hypertarget{repository-creation}{%
\subsubsection{Repository Creation}\label{repository-creation}}

After submitting an issue requesting a package repository as described
above, the repository will be created after review by the \contrib
maintainers. This review does not require that the package actually
exist at this point, but rather is simply intended to check the package
name is reasonable and that the package belongs in \contrib, given
the package description. Initially, the repository will be created as a
private repository until the package authors are ready for the first
release. After the creation of the repository, two or more owners will
be assigned to the repository:

\begin{enumerate}
\def\labelenumi{\arabic{enumi}.}
\tightlist
\item
  The package author requesting the repository.
\item
  The \contrib maintainers.
\end{enumerate}

The \contrib maintainers will not make any changes to the
repository without the express permission of the requesting package
author, and are included in the case of orphaned packages or when
package authors grant permission for minor maintenance to be performed
on the package. The goal is that package authors will have complete
autonomy over their package repository, while still allowing the
\contrib maintainers to update the system as needed. After a
repository is created, the requesting package author can add additional
package authors to the repository via their corresponding
\vb{gitlab.com} accounts. The package authors can
then begin committing to their repository. Note, GitLab includes a whole
suite of features including CI/CD, project wikis, issue trackers,
\emph{etc.} There are a number of GitLab introductions, and GitLab
itself provides
\href{https://docs.gitlab.com/ee/tutorials/gitlab_navigation.html}{introductory
tutorials}.

\hypertarget{version-tagging}{%
\subsubsection{Version Tagging}\label{version-tagging}}

Once a repository is created, the package authors can then develop their
code as they see fit. However, for the \vb{enable}
script of \contrib to be able to interact with the package
repository, the repository must be made public, and versions of the
package must be identified via \vb{git} tags. This
is done as follows:

\begin{lstlisting}[language=bash,style=shell]
git tag VERSION
\end{lstlisting}

Here, \vb{VERSION} is the name of the version that
is being tagged, \emph{e.g.}, \vb{v8318.01}. It is
also possible to create an annotated tag,

\begin{lstlisting}[language=bash,style=shell]
git tag -a VERSION -m 'MESSAGE'
\end{lstlisting}

where \vb{MESSAGE} is now a message associated with
the tag. Once a tag has been created locally, it must be pushed back to
the remote repository on \vb{gitlab.com}.

\begin{lstlisting}[language=bash,style=shell]
git push origin --tags
\end{lstlisting}

It is also possible to push a single tag to the remote repository:

\begin{lstlisting}[language=bash,style=shell]
git push origin VERSION
\end{lstlisting}

Tags can be very useful for any number of reasons, and so the
\contrib system tries to be minimally invasive in how tags are
used for versioning. All version tags \textbf{must} begin with
\vb{v[0-9]}, \emph{i.e.}, the letter
\vb{v} followed by a digit. The remainder of the
version tag can then take on whatever form the package authors wish. Any
tag that does not follow this convention will not be interpreted by
\vb{enable} from \contrib as a version tag.
In this way, package authors can use tags that are not included as
releases. It is important to note that the
\vb{enable} script does not require a version
number to be passed for a package. When this occurs,
\vb{enable} sorts all the
\vb{v[0-9]} tags from the repository by version (as
with \vb{sort -V}) and takes the highest tag as the
most recent version. Care must be taken with versioning schemes, as the
version ordering treats the numbers in a tag as numbers rather than
text, \emph{e.g.}, \vb{v10} is more recent than
\vb{v9}.

While no requirements are made on the version tag format, other than
beginning with \vb{v[0-9]}, the following
versioning conventions are recommended:

\begin{itemize}
\tightlist
\item
  Use tags of the form \vb{v83XX.YY}.
\item
  The \vb{83XX} specifies the \pythia version the
  package was developed for. Note that this does not imply that the
  package is not compatible with other \pythia versions.
\item
  The \vb{YY} specifies the zero-padded version of
  the package for that \pythia version, or alternatively, just specifies
  the package version.
\end{itemize}

Following this convention, package authors can develop separate versions
of a package which are compatible with different versions of \pythia, and
users will understand at a glance which version of \pythia that version
of a package was developed for. Additionally, the most recent version
will be interpreted by the \vb{enable} script as
the tag with the largest package version number for the most recent
version of \pythia, \emph{e.g.}, given the tags
\vb{v8311.01} and
\vb{v8310.06} the most recent version would be
taken as \vb{v8311.01} and not
\vb{v8310.06}.

\hypertarget{testing}{%
\subsubsection{Testing}\label{testing}}

Extensive testing is an important part of any package development. For
\pythia, roughly fifty tests are run before a development branch is
merged, including running all the tests with checks for floating-point
exceptions and memory leaks as well as using a variety of compilers.
This is done through the automated
\href{https://docs.gitlab.com/ee/ci/}{continuous-integration/continuous-deployment
(CI/CD) system} provided by \vb{gitlab.com}. These
tests are performed using \href{https://www.docker.com/}{Docker}
containers. A number of \pythia Docker containers are available through
\href{https://hub.docker.com/u/pythia8/}{Docker Hub}. For development,
the container \vb{pythia8/dev:test} is typically
used, which contains a number of external packages. \pythia can be built
with this container as follows.

\begin{lstlisting}[language=bash,style=shell]
# Pull the Pythia source.
git clone git@gitlab.com:Pythia8/releases.git
cd releases

# Start the Docker container (the cap flag allows use of GDB).
docker run -i -t -v "$PWD:$PWD" -w $PWD -u `id -u` \
  --cap-add=SYS_PTRACE --rm pythia8/dev:test bash --norc

# Configure Pythia with all packages.
export PATH=$PATH:/hep/fastjet3/bin:/hep/hepmc3/bin:/hep/lhapdf5/bin
export PATH=$PATH:/hep/rivet/bin:/hep/yoda/bin
./configure --with-evtgen=/hep/evtgen --with-fastjet3 \
  --with-hepmc2=/hep/hepmc2 --with-hepmc3 --with-lhapdf5 \
  --with-lhapdf6=/hep/lhapdf6 --with-powheg-bin=/hep/powheg/ \
  --with-rivet --with-root --with-gzip --with-python --with-mg5mes \
  --with-openmp --with-highfive=/hep/highfive --with-hdf5=/hep/mpich/ \
  --with-mpich=/hep/mpich/

# Build Pythia.
make -j6

# Build and run the examples.
cd examples
./runmains
\end{lstlisting}

This container, or perhaps a more lightweight container, could be used
for package testing, depending upon the preferences of the package
authors.

Because testing is integral to package development, this is left to the
package authors and not the \contrib maintainers. The
\contrib maintainers will assume all versions provided by package
authors have also been tested and validated by the package authors.
However, if it becomes clear from issues submitted by users that a
specific package is broken, the \contrib maintainers may choose to
exclude this package from the \vb{enable} script
until the latest version of the package is fixed.

Tests run within a package repository typically only call
\vb{make all} within the package. Before publishing
a package, the package build should be tested with the
\href{https://gitlab.com/pythia8-contrib/use}{\vb{use}}
installation of \vb{enable},
\vb{configure}, \vb{make}, and
\vb{make install}. This installation calls
\vb{make contrib} within the package, as described
in the \protect\hyperlink{make-requirements}{make requirements} section,
and assumes the package sits directly below the
\vb{use} top-level directory.

Prior to the release of a new \pythia version, the internal \pythia
developers will provide package authors with a \pythia release candidate.
Ideally, changes to the plugin interfaces between versions will be
minimal, and so package authors should not need to make major changes
between \pythia releases, other than updating the
\vb{PYTHIA8\_PLUGIN\_VERSIONS} macro. The minimal
testing functionality available via the optional
\vb{test} key of
\vb{.meta.yaml}, see the
\protect\hyperlink{documentation}{documentation} section for more
details, is intended to partially automate these checks. The
\contrib maintainers will run these tests prior to a \pythia
release, and notify package authors whether their package is compatible
with the upcoming \pythia release. The results of these tests will be
made available to package authors via the CI of the private
\contrib
\href{https://gitlab.com/pythia8-contrib/website}{\vb{website}}
repository. This automatic testing functionality is currently not
stable, and may change in the future.

\hypertarget{index-generation}{%
\subsubsection{Index Generation}\label{index-generation}}

An index of all the available \contrib packages is available at
\href{https://contrib.pythia.org}{\vb{contrib.pythia.org}}.
For each package, the index is generated as follows:

\begin{itemize}
\tightlist
\item
  name: taken from the package repository.
\item
  description, authors, license, keywords, and dependencies: taken from
  the \vb{.meta.yaml} file of the package, if
  provided, as described in the
  \protect\hyperlink{documentation}{documentation} section.
\item
  documentation: the \vb{README.md} of the package,
  at the version being displayed.
\item
  available versions and compatibility: taken from tagged versions. The
  \pythia compatibility per version is determined automatically from the
  \vb{PYTHIA8\_PLUGIN\_VERSIONS} macro.
\end{itemize}

\hypertarget{advanced-usage}{%
\subsection{Advanced Usage}\label{advanced-usage}}

In this section we provide additional documentation for topics that in
general lie outside the intended scope of the \contrib development
system, but that package authors may nevertheless find relevant.

\hypertarget{additional-classes}{%
\subsubsection{Additional Classes}\label{additional-classes}}

The primary use case for \contrib is through the available base
classes listed above. However, as mentioned in the
\protect\hyperlink{example}{package generation example}, it is possible
to create a class that uses any class available in \pythia as a base
class. To create a template, the \vb{--nocheck}
option should be passed to \vb{generate}. This type
of plugin class cannot be directly passed to \pythia using the
\vb{Init:plugins} setting, and instead must be
loaded using the \vb{make\_plugin} method, which
was described in the \protect\hyperlink{source}{source implementation}
section. This means that while some custom C++ code will need to be
written by the user to access the plugin, that code itself does not need
to link against the plugin library, since its run-time loading is still
handled by the \pythia plugin system.

If settings are required by the plugin library, then before calling the
\vb{make\_plugin} method, the settings of the
plugin library need to be registered with a
\vb{Settings} object. This can be done by the
following method of the \vb{Settings} class.

\begin{lstlisting}[language={C++}]
bool Settings::registerPluginLibrary(string PACKAGE, string INDEX = "")
\end{lstlisting}

Here, \vb{PACKAGE} is the name of the plugin
library optionally with path, and \vb{INDEX} is the
index file for the XML documentation, again with optional path. If
\vb{INDEX} is not provided, and
\vb{PYTHIA8\_PLUGIN\_XML} is defined in the plugin
library, then this location will be used. In this way it is possible to
override the default XML documentation location for the plugin library.
If \vb{PYTHIA8\_PLUGIN\_SETTINGS} is defined within
the plugin library, then the registered settings function will also be
called.

For convenience, the \vb{registerPluginLibrary}
call has been combined with the basic
\vb{make\_plugin} methods into the following
\vb{make\_plugin} method:

\begin{lstlisting}[language={C++}]
shared_ptr<BASE> make_plugin<BASE>(string PACKAGE, string NAME, Pythia* pythiaPtr, const vector<string>& CMNDS)
\end{lstlisting}

This method does the following:

\begin{enumerate}
\def\labelenumi{\arabic{enumi}.}
\item
  Calls \vb{registerPluginLibrary} on the
  \vb{Settings} object from the provided
  \vb{pythiaPtr}. At this point, all settings from
  the plugin library will be available.
\item
  Reads in the settings passed by \vb{CMNDS} using
  the \vb{readString} method of the
  \vb{pythiaPtr}.
\item
  Creates an object of type \vb{NAME} using the
  method,

\begin{lstlisting}
 shared_ptr<BASE> make_plugin<BASE>(string PACKAGE, string NAME, Pythia* pythiaPtr)
\end{lstlisting}

  and then returns a shared pointer of base class
  \vb{BASE}.
\end{enumerate}

In this way it is possible to create with a single method an object from
a plugin library, while still passing valid settings to be used in the
creation of that object. A similar method is also available where the
settings are passed via a file rather than a vector of strings.

\begin{lstlisting}[language={C++}]
shared_ptr<BASE> make_plugin<BASE>(string PACKAGE, string NAME, Pythia* pythiaPtr, string FILE, int SUBRUN = SUBRUNDEFAULT)
\end{lstlisting}

Here, \vb{FILE} is now the settings file and
\vb{SUBRUN} is the subrun to use from that file,
where all settings are read if subruns are not being used and
\vb{SUBRUN} is not passed as an argument. When
using \vb{generate} with the
\vb{--nocheck} argument and an unknown base class,
a test example
\vb{share/PACKAGE/examples/testNAME.cc} is created,
which uses these two methods to create an instance of the plugin class.

It is interesting to note that with this method it is possible to create
a custom class that inherits from the full \pythia class, and then load
it using the plugin system.

\begin{lstlisting}[language=bash,style=shell]
./generate --nocheck MyRecursion Pythia/MyPythia
\end{lstlisting}

\begin{lstlisting}[language={C++}]
Pythia pythia;
vector<string> cmnds = {};
shared_ptr<Pythia> plugin = make_plugin<Pythia>("libMyRecursion.so", "MyPythia", &pythia, cmnds);
\end{lstlisting}

In this way, fully customized versions of \pythia can be provided through
the \contrib package.

\hypertarget{python-packages}{%
\subsubsection{Python Packages}\label{python-packages}}

The \contrib system does not yet support Python modules. However,
the \pythia Python bindings are bidirectional; classes defined in Python
that inherit from C++ classes can be passed back as C++ objects to
\pythia. For example, it is possible to create a
\vb{UserHooks} class entirely in Python.

\begin{lstlisting}[language=Python]
#==========================================================================

# Write own derived UserHooks class.

class MyUserHooks(pythia8.UserHooks):

    # Make sure to initialize base class.
    def __init__(self):
        pythia8.UserHooks.__init__(self)

    # Allow cross section to be modified.
    def canModifySigma(self): return True

    # Double the cross section.
    def multiplySigmaBy(self, sigmaProcessPtr, phaseSpacePtr, inEvent): return 2.
\end{lstlisting}

Here, the class simply doubles the cross section. This can then be
passed back to a \pythia object in Python code as follows:

\begin{lstlisting}[language=Python]
# Generator.
pythia = pythia8.Pythia()

# Pass the user hook back to Pythia.
myUserHooks = MyUserHooks()
pythia.setUserHooksPtr(myUserHooks)
\end{lstlisting}

Note that it is not yet possible to load a class from a Python module,
\emph{i.e.}, write a derived class in Python and then load that class in
C++ code using the \pythia plugin system. However, this may be possible
in the future using Python's ability to be embedded in C.

\hypertarget{library-paths}{%
\subsubsection{Library Paths}\label{library-paths}}

The \pythia and \contrib systems are designed to work
out-of-the-box without user manipulation of library paths, \emph{i.e.},
fixing the error
\vb{cannot open shared object file} reported by
\pythia, even when the plugin library is in a valid location. When this
occurs, there are a number of possible resolutions.

\begin{enumerate}
\def\labelenumi{\arabic{enumi}.}
\item
  Set \vb{LD\_LIBRARY\_PATH} to include the folder
  containing the plugin library, \emph{e.g.},
  \vb{LD\_LIBRARY\_PATH=$LD\_LIBRARY\_PATH:PACKAGE\_PATH}
  where \vb{PACKAGE\_PATH} is the relative or
  absolute path to the plugin library folder. Note,
  \vb{LD\_LIBRARY\_PATH} is not supported in macOS:
  use instead the variable
  \vb{DYLD\_LIBRARY\_PATH}.
\item
  Link or move the plugin library into the same directory as the shared
  \pythia library, \vb{libpythia8.so} or
  \vb{libpythia8.dylib}.
\item
  Link or move the plugin library into the same directory as the
  executable.
\item
  Recompile \pythia with the flag
  \vb{-Wl,--disable-new-dtags} passed to the linker
  via the compiler. This can be done by modifying the following line of
  \vb{Makefile.inc} below. If \pythia 8.312 or later
  is being used, this should be done automatically, yielding:

\begin{lstlisting}
 CXX_SHARED=-shared -Wl,--disable-new-dtags
\end{lstlisting}

  Details on why this flag needs to be passed are given below.
\end{enumerate}

When debugging library path issues, it can be helpful to have further
knowledge of the library path resolution. The
\vb{LD\_DEBUG} variable can be set as follows, and
a full report of the library path resolution will be given.

\begin{lstlisting}[language=bash,style=shell]
LD_DEBUG=libs ./EXE
\end{lstlisting}

Here, \vb{EXE} is the executable being run. Example
output from this, assuming the \vb{MyPackage}
example, might look something like the following, depending on the
system and local development area:

\begin{lstlisting}
     16201: find library=libMyPackage.so [0]; searching
     16201:  search path=../lib/tls/x86_64:../lib/tls:../lib/x86_64:../lib:/home/contrib/develop/releases/lib       (RPATH from file /home/contrib/develop/releases/lib/libpythia8.so)
     16201:   trying file=../lib/tls/x86_64/libMyPackage.so
     16201:   trying file=../lib/tls/libMyPackage.so
     16201:   trying file=../lib/x86_64/libMyPackage.so
     16201:   trying file=../lib/libMyPackage.so
     16201:   trying file=/home/contrib/develop/releases/lib/libMyPackage.so
     16201:  search path=../../../lib/tls/x86_64:../../../lib/tls:../../../lib/x86_64:../../../lib:/home/contrib/develop/releases/lib       (RPATH from file ./testPlugins)
     16201:   trying file=../../../lib/tls/x86_64/libMyPackage.so
     16201:   trying file=../../../lib/tls/libMyPackage.so
     16201:   trying file=../../../lib/x86_64/libMyPackage.so
     16201:   trying file=../../../lib/libMyPackage.so
     16201: 
     16201: 
     16201: calling init: ../../../lib/libMyPackage.so
     16201: 
\end{lstlisting}

From this print-out we see that a number of paths were tried, and then
finally the library was found along the relative path
\vb{../../../lib}. Note that the path
\vb{/home/contrib/develop/releases/lib} was tried
before \vb{../../../lib}. This is because this is
passed as a path to the linker when the shared \pythia library is
compiled; this is also why resolution 2 outlined above should work. In
general, if the plugin library is not being found, this is a very useful
way to see which paths are being tried, and their ordering.

The use of the \vb{disable-new-dtags} flag of
resolution 4 above is necessary because of a change in the default
linker behavior for a number of Linux distributions and the difference
between \vb{RPATH} and
\vb{RUNPATH}. The library path typically resolves
along the following order:

\begin{enumerate}
\def\labelenumi{\arabic{enumi}.}
\tightlist
\item
  \vb{RPATH}: this is the run-time search path that
  is hard-coded into compiled libraries or executables and is typically
  set by passing the flag \vb{-Wl,-rpath,PATH} to
  the linker through the compiler.
\item
  \vb{LD\_LIBRARY\_PATH}: is the environment
  variable used by the dynamic link loader if the library is not found
  along \vb{RPATH}. A similar, but not identical,
  variable \vb{DYLD\_LIBRARY\_PATH} is available
  for macOS.
\item
  \vb{RUNPATH}: this is also passed via the flag
  \vb{-Wl,-rpath,PATH} at compile time, similar to
  \vb{RPATH}.
\end{enumerate}

For older systems, when \vb{rpath} is passed to the
compiler, setting \vb{RPATH} is the default
behavior, while for newer systems \vb{RUNPATH} is
set instead. Consequently, depending on whether
\vb{RPATH} or \vb{RUNPATH} is
preferred, the following options should be used:

\begin{itemize}
\tightlist
\item
  older systems: for \vb{RPATH} no additional flags
  need to be passed, while for \vb{RUNPATH} the
  flag \vb{-Wl,--enable-new-dtags} must be passed
  to the compiler.
\item
  newer systems: for \vb{RPATH} the flag
  \vb{-Wl,--disable-new-dtags} must be passed to
  the compiler, while for \vb{RUNPATH} no
  additional flags need to be passed.
\end{itemize}

For some systems, it is possible that the
\vb{disable-new-dtags} and
\vb{enable-new-dtags} options are not available, in
which case the \vb{rpath} flag will typically set
\vb{RPATH} and not
\vb{RUNPATH}. Note that it is also possible to
avoid passing \vb{rpath} flags to the linker by
instead setting the \vb{LD\_RUN\_PATH} environment
variable before running the linker.

There is a critical difference between \vb{RPATH}
and \vb{RUNPATH}: \vb{RPATH}
is transitive while \vb{RUNPATH} is not. In other
words, the \vb{RPATH} for a dependency will be used
in addition to the \vb{RPATH} for the binary
itself. Conversely, if \vb{RUNPATH} is specified,
then only that \vb{RUNPATH} for that binary will be
used, even if a dependency specifies an \vb{RPATH}.
From the Linux Foundation
\href{https://refspecs.linuxfoundation.org/elf/gabi4+/ch5.dynamic.html\#shobj_dependencies}{reference
specifications}, the following is noted about the lack of the transitive
property for \vb{RUNPATH}:

\begin{quote}
The set of directories specified by a given
\vb{DT\_RUNPATH} entry is used to find only the
immediate dependencies of the executable or shared object containing the
\vb{DT\_RUNPATH} entry. That is, it is used only
for those dependencies contained in the
\vb{DT\_NEEDED} entries of the dynamic structure
containing the \vb{DT\_RUNPATH} entry, itself. One
object's \vb{DT\_RUNPATH} entry does not affect the
search for any other object's dependencies.
\end{quote}

This has important implications for \contrib. If \pythia is
compiled with \vb{RUNPATH}, then even if
\vb{RPATH} is correctly specified for the
executable calling \pythia, \pythia will not use this path at run time.
Similarly, if the executable is compiled with
\vb{RUNPATH} then \pythia will not search along this
path. Consequently, for \pythia to search along a path specified by the
executable, both \pythia and the executable must be compiled using
\vb{RPATH} and not
\vb{RUNPATH}.